\documentclass{aa}
\usepackage[varg]{txfonts}

\usepackage[colorlinks=true,     linkcolor=blue, citecolor=blue, filecolor=blue, urlcolor=blue]{hyperref}
\usepackage{graphicx,rotating}
\usepackage{amssymb,amsmath,pdflscape}
\usepackage{natbib}
\usepackage{empheq}
\usepackage{upgreek}

\newcommand{\hatrho}{{\hat{\rho}}}
\newcommand{\hatomega}{{\hat{\varOmega}}}
\newcommand{\du}{{\rm d}}
\newcommand{\rp}{R_{\rm p}}
\newcommand{\re}{R_{\rm e}}
\newcommand{\const}{C^{\rm te}}
\newcommand{\bare}{\bar{\varepsilon}}

\begin{document}

\title{Approaching the structure of rotating bodies\\ from dimension reduction\thanks{A minimum driver program is available at \url{https://github.com/clstaelen/ssba}}}

\titlerunning{The structure of rotating bodies from dimension reduction}

\author{C. Staelen \and J.-M. Hur\'e}
\institute{Univ. Bordeaux, CNRS, LAB, UMR 5804, F-33600 Pessac, France \\ \email{clement.staelen@u-bordeaux.fr}}

\date{Received ??? / Accepted ???}

\abstract{
    We show that the two-dimensional structure of a rigidly rotating self-gravitating body is accessible with relatively good precision by assuming a purely spheroidal stratification. With this hypothesis, the two-dimensional problem becomes one-dimensional, and consists in solving two coupled fixed-point equations in terms of equatorial mass density and eccentricity of isopycnics. We propose a simple algorithm of resolution based on the self-consistent field method. Compared to the full unconstrained-surface two-dimensional problem, the precision in the normalized enthalpy field is better than $10^{-3}$ in absolute, and the computing time is drastically reduced. In addition, this one-dimensional approach is fully appropriate to fast rotators,  works for any density profile (including any barotropic equation of state), and  can account for mass density jumps in the system, including the existence of an ambient pressure. Several tests are given.}

\keywords{Gravitation - stars: interiors - stars: rotation - planets and satellites: interiors - Methods: numerical}

\maketitle

\section{Introduction}

The structure of static polytropic stars,  in the classical sense,  is traditionnally described by the Lane-Emden equation, which admits a wide variety of solutions  \citep[e.g.,][]{sri62,sha77,sei78,liu96,horedttextbook2004,mach12,tohlinewiki21}. With rotation, the shape and the internal isobars deviate from sphericity, and it becomes difficult to anticipate precisely the topology of the gravitational field lines and to make direct use of the Gauss theorem. The context of slow rotation is attractive as it enables us  to perform various kinds of analytical expansions,  for instance in the form of series \citep{ch33,cr63,rs66,kov68}. Unfortunately, the treatment required for moderate and fast rotations is much harder; there is almost no analytical way \citep{rob63}. The Poisson equation must be solved numerically, while the fluid boundary is not known in advance. Obviously, numerical methods offer a better range of options; for instance, they can model almost any kind of rotation profile, flattening, and equation of state (EOS) \citep[e.g.,][]{j64,om68,hachisu86}; they  include magnetic fields \cite[e.g.,][]{te05,lj09}, ambient pressure \cite[e.g.,][]{hhn18}, mass density jumps \citep{kiu10,ka16,bh21};  and they can reach three-dimensional configurations and multiplicity \citep{hachisu86III,et09}. In this last context, the self-consistent field (SCF) method of resolution, which consists in finding a fixed point for the mass density $\rho(\vec{r})$ from the pertinent equation set, is very efficient when appropriately initialized and scaled, and it has largely been used to model stars, binaries, and rings \citep{hachisu86,od03}. However, it is known from classical theories that, at slow rotations, the equilibrium configurations remain very close to ellipsoids of revolution \citep[e.g.,][]{veronet12,ch33}, and take a slightly  sub-elliptical shape in a meridian plane   between the pole and the equator \citep[see also, e.g.,][]{cmm15}. This assessment also applies for fast rotations, except close to the mass-shedding limit \citep{sh24}. 

Any assumption made upon the mass density structure or on the symmetry is expected to reduce the mathematical complexity of the problem, but it diverts from the exact problem. This is the case when isobaric or isopycnic surfaces are locked to spheroids (i.e., ellipses in the meridional plane). Actually, under axial symmetry it is possible to  benefit from Newton's and Maclaurin's theories and to use the closed-form for the potential of the homogeneous spheroid \citep{chandra69,binneytremaine87}, and subsequently construct heterogeneous bodies by piling-up coaxial homogeneous spheroids \citep[see, e.g.,][]{ak74,mmc83}. In a series of papers \cite[][hereafter Papers I, II, and IV, respectively]{h2022a,h2022b,sh24}, we   developed a theory that solves the equilibrium of a heterogeneous body made of ${\cal L}$ homogeneous layers bounded by spheroids with different eccentricities, and in asynchronous rotational motion. As shown, it is possible to determine a relationship between the rotation rate $\Omega_i$ of each layer $i \in [1,{\cal L}]$, the parameter set of the spheroids ${\rm E}_i$, and the mass-densities $\rho_i$ involved. This theory is approximate, but works very well provided the interfaces between layers are close enough to be confocal with each other. In the limit ${\cal L}\rightarrow \infty$, we   showed that, for global rigid rotation, the eccentricity ${\varepsilon=[1-b^2/a^2]^{1/2}}$ of isopycnic surfaces (with semimajor axis $a$ and semiminor axis $b$) and the mass density profile $\rho$ obey a general integro-differential equation (IDE) (see Eq. (19) in Paper IV and Eq. \eqref{eq:id_nsfoe} below). This IDE works very well for a wide range of rotation rates (or flattenings). It encompasses Clairaut's fundamental equation in the limit of small flattenings, and behaves correctly even close to the mass-shedding limit. This is important in order to provide tools appropriate to fast rotators, which may not be rare entities in the Universe \citep{ramp23}.

The assumption of spheroidal isopycnics is undoubtedly strong, but it is not only  motivated     by mathematical simplifications coming out. It is widely supported by the numerical experiments. Actually, the true surface of a rigidly rotating fluid is generally very close to spheroids for a wide range of flattenings, even in the presence of mass density jumps. For instance, for a polytropic gas with index $n=1$ and a polar-to-equatorial axis ratio $\rp/\re=0.95$, the shapes of isopycnics deviate in altitude from ellipses by less than $10^{-3}$ in relative from the center to the surface; the volumes differ by a similar value. For some problems, the deviations are  too large and the hypothesis of spheroidal stratification must be abandoned \citep[e.g.,][]{zt70,hub13,cmm15,net17,dc18}. However, there are many situations where, on the contrary, such a precision is sufficient, for instance for the construction of mass-radius relationships or for dynamical studies  \citep[e.g.,][]{had00,kr13,mv15,vdap20}.

In this article we propose a direct exploitation of the IDE derived in Paper IV, looking for self-consistent solutions. As the underlying equation set is reduced to the Bernoulli equation combined with the Poisson equation, the solution  necessarily has a limited range of applications, as quoted above. Our approach  cannot compete with sophisticated models for the stellar structures for instance \citep[see, e.g.,][for the state of art]{r06,elr13,rep16,hr23}. Our main  aim is to analyze the performance of the spheroidally stratified barotrope (SSB) approximation, which to our knowledge has never been reported. By construction, the IDE concerns only equatorial values. When it is combined with the centrifugal equilibrium along the polar axis, we obtain a type of equatorial projection of the original bidimensional equation set. This opens the possibility to determine the two-dimensional structure of a fully heterogeneous rigidly rotating body from a one-dimensional approach. This is explained in Sect. \ref{sec:theory}. An attractive point is that the IDE is valuable regardless of the EOS, leaving a certain flexibility in terms of applications. We discuss in Sect. \ref{sec:bmscf} a simple iterative algorithm to solve this set of projected equations. The cycle is based on the classical self-consistent field (SCF) method. In the present case, we have to solve two coupled fixed-point problems, one for the eccentricity profile $\varepsilon(a)$ and one for the mass density profile $\rho(a)$, in the range $a \in [0,\re]$. The full $2$D structure is   reconstructed   by unfolding the equatorial solution $\{\rho(a),\varepsilon(a)\}$. Several tests are proposed in Sect. \ref{sec:tests}, including static, slowly rotating, and highly rotating configurations, as well as Hachisu's equilibrium sequences \citep{hachisu86}. For this purpose, a numerical reference is needed to compare the approximate solution to the real one. We use the {\tt DROP} code, which computes the equilibrium configurations in full $2$D \citep{hh17,bh21}, the Poisson being solved by the multigrid-method from finite-difference equation. Using the spectral version of the code would not help much as we are dealing with deviations that are not tiny, typically on the order of $10^{-4}$. In Sect. \ref{sec:jumps} we show how to account for mass density jumps, which enables us to model multi-domain bodies. Two examples are discussed, namely the pressurized $n=5$ polytrope and the Earth's interior, by using the nonrotating Preliminary Reference Earth Model \citep[][]{prem81}. A summary is found in the last section.

\section{Theoretical background}
\label{sec:theory}

Basically, a star is an equilibrium state between gas pressure, gravity forces, and centrifugation. In the barotropic approximation, the relevant equation set is

\begin{subequations}\label{eq:eq_set}
    \begin{empheq}[left={\empheqlbrace\,}]{align}
        &H +\varPhi + \varPsi = \const,\label{eq:bernoulli}\\
        &\nabla^2\varPsi = 4\uppi G\rho,\label{eq:poisson}\\
        &f(H,\rho)=0\label{eq:eos},
    \end{empheq}
\end{subequations}
where $\varPsi$ is the gravitational potential, $\varPhi = - \int{\du R\,\varOmega^2 R}$ is the centrifugal potential, $\varOmega$ is the local rotation rate, $P$ is the pressure, $\rho$ is mass density of the fluid, and $H= \int\du P/\rho$ is the enthalpy (the self-gravitating flow is isentropic). The link between $H$ and $\rho$ in Eq.\eqref{eq:eos} is usually done via the EOS, namely $P(\rho)$. This set represents a conservative form for the flow, is independent of time, ignores viscosity, and supposes that rotation is constant along cylinders, coaxial with the rotation axis \citep[e.g.,][]{tassoul78,amendt1989}, which means that  $\varPhi$ depends on the cylindrical radius $R$ only. In the present article we consider axially symmetric configurations in the framework of the nested spheroidal figure of equilibrium (NSFoE) reported in Papers I and II, which assumes that, in a layered system, all the surfaces bounding the homogeneous layers (${\cal L}$ in total, $\rho_i$ is the mass density of layer $i$) are perfect spheroids ${\rm E}_i(a_i,b_i)$. In the meridional plane, these surfaces are ellipses with equation
\begin{equation}
  \frac{R^2}{a_i^2}+\frac{Z^2}{b_i^2}=1,
  \label{eq:ellipse}
\end{equation}
where $a_i \in [0,\re]$, $b_i \in [0,R_{\rm p}]$ and $\re$ and $R_{\rm p}$ are the equatorial and polar radii of the body, respectively. In these conditions, the equilibrium (if it exists) is perfectly determined as a function of the set $\{\rho_i, {\rm E}_i\}_{i\in[1,{\cal L}]}$.

\subsection{Dimension reduction (equatorial projection)}

In Paper IV we   showed that, in the limit where the number ${\cal L}$ of layers is infinite, Eqs. \eqref{eq:bernoulli} and \eqref{eq:poisson}
can be formally solved regardless of any EOS. In the special case of rigid rotation   considered here, we obtain an IDE linking the eccentricity profile $\varepsilon(a)$ and the mass density profile $\rho(a)$. In order to render the problem scale-free, the semimajor axis $a$ of a given isopycnic surface is expressed in units of the equatorial radius $\varpi = a/R_{\rm e}\in[0,1]$, and  $\hat\rho=\rho/\rho_{\rm c}$ is the dimensionless mass density ($\rho_{\rm c}$ is the central mass density). In these conditions, the IDE reads
\begin{equation}\label{eq:id_nsfoe}
  \frac{\du\varepsilon^2}{\du \varpi} = \frac{2\int_{\hatrho(0)}^{\hatrho(\varpi)} \du\hatrho(\varpi') \chi(\varpi',\varpi;\varepsilon)}{\int_{\hatrho(0)}^{\hatrho(1)} \du\hatrho(\varpi')\mu(\varpi',\varpi;\varepsilon)},
\end{equation}
where $\chi$ and $\mu$ are continuous smooth functions in the whole domain $\varpi \in [0,1]$. These are defined in Paper IV (see also  Appendix \ref{app:chimunueta}, respectively Eq.\eqref{eq:chi_def} and Eq.\eqref{eq:mu_def}). As explicitly stated, $\chi$ and $\mu$ depend on the radius $\varpi$ and on the eccentricity profile $\varepsilon(\varpi)$. We can safely replace Eqs. \eqref{eq:bernoulli} and   \eqref{eq:poisson} by Eq.  \eqref{eq:id_nsfoe}, and express Eq. \eqref{eq:bernoulli} along the polar axis. This is possible as the gravitational potential is known. As the constant in the RHS, we can take the value of the LHS at the pole, which is the most straightforward. After some algebra (see Appendix \ref{app:hent_proof} for a proof), we find for the dimensionless enthalpy
\begin{align}
    \hat{H}(\varpi) &= \hat{H}(1)+ 2 \uppi\int_{\hat\rho(0)}^{\hat\rho(1)}\mkern-7mu\du\hat{\rho}(\varpi')\left[\eta(\varpi',1;\varepsilon)-\eta(\varpi',\varpi;\varepsilon)\right]\notag,
\end{align}
where $\hat{H}=H/G\rho_{\rm c}R_{\rm e}^2$, and $\eta$ is reported in    Appendix \ref{app:chimunueta} (see  Eq.\eqref{eq:eta_def}). It follows that Eq.\eqref{eq:eq_set}, in the framework of the theory of NSFoE, becomes
\begin{subequations}\label{eq:eq_set_1D}
    \begin{empheq}[left={\empheqlbrace\,}]{align}
        & \hat{H}(\varpi) = \hat{H}(1)\label{eq:hent}\\
        & \qquad + 2 \uppi\int_{\hat\rho(0)}^{\hat\rho(1)}\mkern-7mu\du\hat{\rho}(\varpi')\left[\eta(\varpi',1;\varepsilon)-\eta(\varpi',\varpi;\varepsilon)\right] \nonumber\\
      &\frac{\du\varepsilon^2}{\du\varpi} = \frac{2\int_{\hat\rho(0)}^{\hat\rho(\varpi)} \du\hat\rho(\varpi')\chi(\varpi',\varpi;\varepsilon)}{\int_{\hat\rho(0)}^{\hat\rho(1)} \du\hat\rho(\varpi')\mu(\varpi',\varpi;\varepsilon)},\label{eq:ide_compact}\\
      &  f\big(\hat{H}(\varpi),\hat\rho(\varpi)\big)=0\label{eq:eos_equat},
    \end{empheq}
\end{subequations}
where the last equation is the EOS. We note that, if the mass density profile happens to be known in advance (e.g., deduced from observational data) or prescribed, Eq.\eqref{eq:hent} is not needed anymore (see, e.g., Sect. \ref{subsec:earth}). We immediately see that the problem is now fully one-dimensional. It means that we can compute the $2$D structure of a rotating gaseous body from a $1$D approach. This dimension reduction thus relies on a SSB approximation.

\begin{figure}
    \centering
    \includegraphics[width=0.75\linewidth,trim={0.cm 0.cm 0.cm 0.cm},clip]{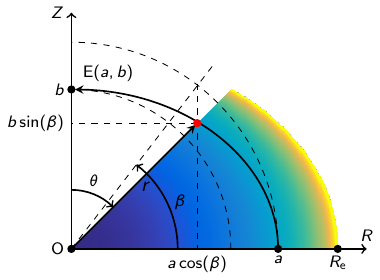}
    \caption{Schematic of the unfolding process. The one-dimensional equation set is first solved for the mass density $\rho$ and eccentricity $\varepsilon$ along the equatorial axis in the full interval $[0,\re]$. Then, the two-dimensional structure is reconstructed by propagating, for each radius $(a,0)$, the solution (enthalpy, mass density, or pressure) along the ellipse ${\rm E}(a,b)$ up to the polar axis at $(0,b)$ (see Sect. \ref{subsec:ufolding}).}
    \label{fig:reconstruction}
\end{figure}

\subsection{Unfolding and global quantities}
\label{subsec:ufolding}

Once the $1$D solution $\{\hat\rho(\varpi),\varepsilon(\varpi)\}$ is calculated, we can reconstruct any bidimensional map by unfolding the mass density, at each radius $a$, along the corresponding ellipse ${\rm E}(a,b)$ up to the polar axis at $(0,b)$, as depicted in Fig. \ref{fig:reconstruction}. In spherical polar coordinates $(r,\theta,\varphi)$, we actually have
\begin{equation}\label{eq:unfolding}
\hatrho(r,\theta)=\hatrho(\varpi),
\end{equation}
where the two coordinates are easily determined by using the parametric formula
\begin{subequations}\label{eq:parametric}
  \begin{empheq}[left={\empheqlbrace\,}]{align}
    &r^2=\re^2\varpi^2\left[1 - \varepsilon^2(\varpi) \sin^2 (\beta)\right],\\
    &\tan(\theta) = \frac{1}{\bare(\varpi)}\cot(\beta),
    \end{empheq}
\end{subequations}
with $\theta \in [0,\uppi/2]$ and $\beta \in [0,\uppi/2],$ and where  
\begin{equation}\label{eq:bare_def}
    \bare(\varpi) = [1-\varepsilon^2(\varpi)]^{1/2}
\end{equation}
is the axis ratio of the isopycnic ${\rm E}(a,b)$ considered. The conversion to cylindrical coordinates is straightforward: ${R= \re\varpi\cos(\beta)}$ and ${Z = \re\varpi\bare(\varpi)\sin (\beta)}$. We can subsequently deduce all global quantities (e.g., mass, volume). 

In $(\varpi,\beta,\varphi)$ coordinates, the volume element writes
\begin{equation}
    \frac{\du V}{\re^3} = 2\uppi\frac{\varpi^2\cos(\beta)}{\bare(\varpi)}\left[\bare^2(\varpi)-\frac12\sin^2(\beta)\frac{\du\varepsilon^2}{\du\varpi}\right]\du\varpi\du\beta\du\varphi.
\end{equation}
For most global properties of the system, the integrations over $\beta$ are trivial. For instance, the mass $M$ and the moment of inertia $I$ are reduced to a single integral over $\varpi$ and we have
\begin{equation}
    \left\{
        \begin{aligned}
            &\frac{M}{\rho_{\rm c}\re^3} = 4\uppi\int_0^1\du\varpi\frac{\varpi^2\hatrho(\varpi)}{\bare(\varpi)}\left[\bare^2(\varpi)-\frac16\frac{\du\varepsilon^2}{\du\varpi}\right],\\
            &\frac{I}{\rho_{\rm c}\re^5} = \frac83\uppi\int_0^1\du\varpi\frac{\varpi^4\hatrho(\varpi)}{\bare(\varpi)}\left[\bare^2(\varpi)-\frac1{10}\frac{\du\varepsilon^2}{\du\varpi}\right].
        \end{aligned}
    \right.
\end{equation}

According to Paper IV, the rotation rate $\varOmega$ is given by
\begin{equation}
    \hatomega^2 = - 2\uppi\int_{\hatrho(0)}^{\hatrho(1)} \du\hatrho(\varpi')\kappa(\varpi',\varpi;\varepsilon),\label{eq:rrate}
\end{equation}
where $\hatomega = \varOmega/\sqrt{G\rho_{\rm c}}$ and $\kappa$ is defined by Eq. \eqref{eq:kappa_def} (see again Paper IV). It is worth recalling that, as the theory of NSFoE is an approximate theory, Eq.\eqref{eq:rrate} is expected to exhibit slight variations with the radius (see below), and should rigorously be denoted $\hatomega^2(\varpi)$.

\subsection{Comments}

As can be seen in Paper IV, the three functions $\chi$, $\mu$, and $\eta$ have small amplitude, but all take real, negative values in the range of $\varpi$ of interest. It is therefore clear that, if the integral in Eq.\eqref{eq:hent} happens to be essentially negative, then the mass density can become negative, in which case the computed solution cannot be compatible with a physical solution. Density inversions are eventually acceptable, but $\rho \ge 0$ is a firm condition. A similar remark holds for Eq. \eqref{eq:rrate}. For Eq. \eqref{eq:ide_compact}, things are less restrictive. Actually, negative values of $\varepsilon^2$ correspond to prolate spheroids. This is not a problem because there is a mathematical continuity in the gravitational potential when the eccentricity becomes a purely imaginary number  (i.e., when $\varepsilon = 0^+ \rightarrow {\rm i} 0^+$; \citealt{binneytremaine87}). From a numerical point of view,  however, the requires a dedicated treatment, and it seems preferable to consider the axis ratio, namely $\bare(\varpi) \lessgtr 1$ instead of $\varepsilon$ (see below, and  Appendix \ref{app:onefunction}).

\section{Solution with a SCF-algorithm. Implementation and example}
\label{sec:bmscf}

\subsection{Cycle and convergence criterion}

It is well known that Eqs.\eqref{eq:bernoulli} and \eqref{eq:poisson} correspond to a fixed-point problem:  $\rho=f(\rho)$ in terms of the mass density profile. Clearly, the equation set Eq.\eqref{eq:eq_set_1D} has a similar structure, but we have two coupled fixed-point  problems of the form
\begin{subequations}\label{eq:bimodalscf}
    \begin{empheq}[left={\empheqlbrace\,}]{align}
        &\hatrho^{(t)} \leftarrow f_1\left(\hatrho^{(t-1)},\bare^{(t-1)}\right),\label{eq:assignrho}\\
        &\bare^{(t)} \leftarrow f_2\left(\hatrho^{(t-1)},\bare^{(t-1)}\right),\label{eq:assignepsilon}
    \end{empheq}
\end{subequations}

It is therefore natural to proceed in the same manner as for the standard SCF method: we guess the profiles for the mass density and the axis ratio, namely $\hatrho^{(0)}(\varpi)$ and $\bare^{(0)}(\varpi)$, and let the profiles evolve until stabilization. This can be accomplished following the iterative scheme, for $t\ge 1$:
\begin{enumerate}
    \item $\bare^{(t)}$ is obtained from Eq. \eqref{eq:bare_def}, after integrating Eq. \eqref{eq:ide_compact};
    \item $\hat{H}^{(t)}$ is computed according to Eq. \eqref{eq:hent};
    \item $\hatrho^{(t)}$ is obtained from Eq. \eqref{eq:eos_equat}.
\end{enumerate}
We see that there are two other options (with no significant impact on the performance of the cycle), depending on the order of assignments; in other words,  $\bare$ in Eq.\eqref{eq:assignepsilon} can be computed using the $\rho$-profile updated from Eq. \eqref{eq:assignrho}, or $\rho$ in Eq. \eqref{eq:assignrho} can be computed from the $\bare$-profile updated from Eq. \eqref{eq:assignepsilon}. At convergence, two successive profiles must be numerically unchanged, in which case the cycle ends. As we traditionally use double precision computers, we take\footnote{In practice, this level of stability is easily reached. However, it needs to be raised if the numerical resolution is very high.}$^{,}$\footnote{This may appear useless or excessive. In some cases, however, the cycle first converges toward one state and then ``bounces'' to reach a completely different one. Reaching machine precision offers   better confidence in the solution.} $\delta^{(t)} \lesssim 10^{-14}$, where
\begin{equation}\label{eq:delta}
    \delta^{(t)}~\equiv~ \max\left(\Delta^{(t)}\hatrho,\Delta^{(t)} \bare\right)
,\end{equation}
where we define still for $t \ge 1$ and $\varpi \in [0,1]$
\begin{equation}
  \Delta^{(t)} \hatrho = \sup \left| \hatrho^{(t)}(\varpi)-\hatrho^{(t-1)}(\varpi)\right|,
\end{equation}
and  use a similar definition for $\Delta^{(t)} \bare$.

\subsection{Boundary conditions}
In the standard (single-domain) case, and in the absence of external pressure (see Sect. \ref{subsec:extp}), we can take Dirichlet boundary conditions (BCs) at the outer boundary $\varpi=1$
\begin{equation}\label{eq:bcs}
    \left\{
        \begin{aligned}
            &\hat{\rho}(1)=0\\
            &\bare(1)=\bare_{\rm s}
        \end{aligned},
    \right.
\end{equation}
where $\bare_{\rm s}=\rp/\re$ is the axis ratio of the outermost surface.  At the center of the body, we have
\begin{equation}\label{eq:bccenter}
    \left\{
        \begin{aligned}
            &\hat{\rho}(0)=1\\
      &\left.\frac{\du\bare^2}{\du\varpi}\right|_{\varpi=0}=0
        \end{aligned}.
    \right.
\end{equation}
Unfortunately, $\bare(0)$ cannot be easily deduced.

\subsection{Method and implementation (quadrature schemes, \texorpdfstring{$\varpi$}{}-grid, and seeds)}
\label{subsec:implementation}

We see that Eq. \eqref{eq:eq_set_1D} involves derivatives and quadratures, and there are different ways to estimate $\rho(\varpi)$ and $\bare(\varpi)$ for given BCs. Here we decided to recast Eq. \eqref{eq:ide_compact} in integral form, i.e., 
\begin{align}\label{eq:integ_eps}
  \bare^2(\varpi)= \bare_{\rm s}^2-\int_{\varpi}^{1}{\left(\frac{\du\bare^2}{\du\varpi}\right)\du\varpi},
\end{align}
where the term in parentheses is simply the right-hand side of Eq. \eqref{eq:ide_compact} within a minus sign. So, we have only to deal with quadratures. The eccentricity being unknown at the center, we chose to integrate downward, from the surface to the center, with Eq. \eqref{eq:bcs} as Dirichlet BCs. The computational grid is made of $N+1$ equally spaced nodes: $\varpi_i \in \{0,\frac{1}{N},\,\dots,\,\varpi_i=\frac{i}{N},\,\dots,\,1\}$. We used the trapezoidal rule as the quadrature rule, which is second-order in the step size (more efficient schemes can be used at this level). According to Eq.\eqref{eq:bcs}, still assuming null ambient pressure, we take
\begin{subequations}\label{eq:init}
    \begin{empheq}[left={\empheqlbrace\,}]{align}
        &\hat{H}^{(0)}(\varpi) = \hat{H}_{\rm c}(1-\varpi^2),\label{eq:hent_init}\\
        &\bare^{(0)}(\varpi) = 1-(1-\bare_{\rm s})\varpi^2\label{eq:ebar_init}
    \end{empheq}
\end{subequations}
as the starting guess. Initially, we thus have $\hat{H}^{(0)}(0)=\hat{H}_{\rm c}$ at the center, and $\bare^{(0)}(0)=1$. These quadratic profiles seem appropriate for a wide range of configurations, although the solutions   generally have a nonzero central eccentricity. Obviously, the BCs must be applied at each step in the cycle.

\subsection{A note about the equation of state}

The EOS is the fundamental ingredient. Without loss of generality, we consider a polytropic gas where the pressure is a power law  of the mass density, which leads to the expression $H=K(n+1)\rho^{1/n}$, where $K$ is a positive constant and the polytropic index $n>0$ is finite. In this case the relationship between $\hat{H}$ and $\hat\rho$ is 
\begin{equation}
    \frac{\hat{H}}{\hat{H}_{\rm c}}=\hat\rho^{1/n},
\end{equation}
where $\hat{H}_{\rm c} = K(n+1)/G\rho_{\rm c}^{1-1/n}R_{\rm e}^2$ (this relation assumes $\hat\rho\ge 0)$. The mass density $\hatrho^{(t)}$ inside the cycle, including the seed, is deduced from the EOS through Eq. \eqref{eq:eos_equat}.

\begin{figure}
    \centering
    \includegraphics[width=\linewidth,trim={0.2cm 0.2cm 0.2cm 0.2cm},clip]{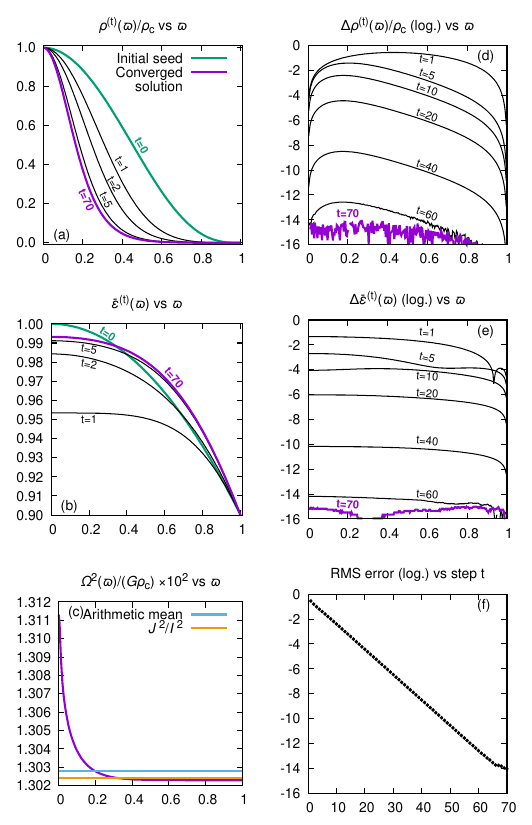}
    \caption{Radial profiles for $\hat\rho$ (panel a) and   $\bare$ (panel b) during the SCF cycle for a few selected steps $t$ (labeled on the curves) of the SCF cycle, and absolute deviations (panels d and e) between steps $t-1$ and $t$. Also shown are the square of the rotation rate (panel c) computed at convergence from Eq. \eqref{eq:rrate} and its mean value (orange), and the convergence parameter $\delta$  computed from Eq. \eqref{eq:delta} as a function of   step $t$ (panel f).}
    \label{fig:ecs_prog}
\end{figure}

\begin{table}[ht]
    \centering
    \caption{Global quantities computed for configuration A discussed in Sect. \ref{subsec:implementation} with the SSB approximation for $N=256$ (last column). The rotation rate is a mean value (see text). Also listed (second column) are values output from the {\tt DROP} code at a resolution $257 \times 257$.}
    \begin{tabular}{lrr}
    configuration A    & {\tt DROP}$^{\dagger}$ code & this work\\\hline    
        $\bare_{\rm s}$ & $\leftarrow 0.900$ & $\leftarrow 0.900$ \\
        $n$ & $\leftarrow 3.0$ & $\leftarrow 3.0$ \\
        $M/(\rho_{\rm c}R_{\rm e}^3)$ & $5.9999\times10^{-2}$ & $6.0062\times10^{-2}$ \\
        $I/(\rho_{\rm c}R_{\rm e}^5)$ & $3.9410\times10^{-3}$ & $3.9475\times10^{-3}$ \\
        $V/R_{\rm e}^3$ & $3.7436\textcolor{white}{xxxxxx}$ & $3.7699\textcolor{white}{xxxxxx}$\\
        $\hatomega^2$ & $1.3006\times10^{-2}$ & $1.3024\times10^{-2}$ \\
                      &                      & $^\star1.3014\times10^{-2}$\\
        $J/(G\rho_{\rm c}^3R_{\rm e}^{10})^{1/2}$ & $4.4945\times10^{-4}$ & $4.5051\times10^{-4}$ \\
        $-W/(G\rho_{\rm c}^{2}R_{\rm e}^{5})$ & $5.8487\times10^{-3}$ & $5.8599\times10^{-3}$ \\
        $T/(G\rho_{\rm c}^{2}R_{\rm e}^{5})$ & $2.5629\times10^{-5}$ & $2.5707\times10^{-5}$ \\
        $U/(G\rho_{\rm c}^{2}R_{\rm e}^{5})$ & $5.7976\times10^{-3}$ & $5.8077\times10^{-3}$\\
        $|{\rm VP}/W|$ & $4\times10^{-5}$ & $1\times10^{-4}$ \\
        \hline
        \multicolumn{3}{l}{$\leftarrow$ input data}\\
        \multicolumn{3}{l}{$\dagger$ see \cite{hh17} and \cite{bh21}}\\
        \multicolumn{3}{l}{$^\star$ value obtained for $N=1024$}\\
    \end{tabular}
    \label{tab:configA}
\end{table}

\subsection{Example of cycle convergence}
\label{subsec:example}
  
The first example is for $\bare_{\rm s}=0.9$ and $n=3$, and the grid has $N+1=257$ equally spaced nodes. This configuration corresponds, for instance, to a radiation-pressure-dominated ideal gas or to a white dwarf with fully degenerate extremely relativistic electrons. With this parameter, we are already beyond slow rotators. We   ran the code based on the SSB approximation. Figure \ref{fig:ecs_prog} shows the evolution of $\hat\rho(\varpi)$ and $\varepsilon^2(\varpi)$ during the cycle, as well as the deviations $\Delta \hatrho$ and $\Delta \varepsilon^2$ from one step to the other. Figure \ref{fig:ecs_prog}f gives the convergence parameter $\delta^{(t)}$ defined by Eq. \eqref{eq:delta} from the beginning to convergence. We see that $\delta^{(t)}$ decreases exponentially with the step $t$ and the algorithm converges quickly on a solution. Convergence is reached after 70 cycles with the current criterion.  After step 20, we   already have $\delta^{(t)}\sim10^{-5}$, which is on the order of the accuracy of the numerical scheme (i.e., $1/N^2 \sim 10^{-5}$ with $257$ nodes). The next steps are necessary to reach the threshold of $10^{-14}$. We show in Fig. \ref{fig:ecs_prog}c the rotation rate   computed from Eq. \eqref{eq:rrate}. Unsurprisingly, there is a slight variation with the radius, on the order of $10^{-2}$ in relative. This is due to the approximate nature of  the theory of NSFoE. The function $\kappa$ involved when computing $\Omega^2$ changes sign for most pairs $(\varpi',\varpi)$, which  in some cases results in subtracting two quantities close to each other, thereby amplifying errors. This effect can be reduced by increasing the numerical reolution. Actually, a test with $N=1024$ shows a mean value $\langle\hat\varOmega^2\rangle\approx1.3014\times10^{-2}$ and an amplitude around $10^{-3}$ in relative, which is more reasonable.

We list in Table \ref{tab:configA} the main global quantities at equilibrium, namely the mass $M$, the moment of inertia $I$, the volume $V$, the angular momentum $J$, the gravitational energy $W$, the kinetic energy $T$, the internal energiy $U$, and the virial parameter ${\rm VP}~=~W+2T+U$. We see that ${\rm VP}/|W|$ is on the order of $10^{-4}$ in absolute, which is very good.
We also list the values output by the {\tt DROP} code, which solves Eq. \eqref{eq:eq_set} in full $2$D \citep{hh17,bh21}. We see that the agreement between the two methods is quite good, with deviations on the order of  a few $10^{-3}$ in relative, while the resolution is moderate.

\begin{figure}
    \centering
    \includegraphics[height=0.6\linewidth,trim={8.2cm 0.2cm 8cm 0.4cm},clip]{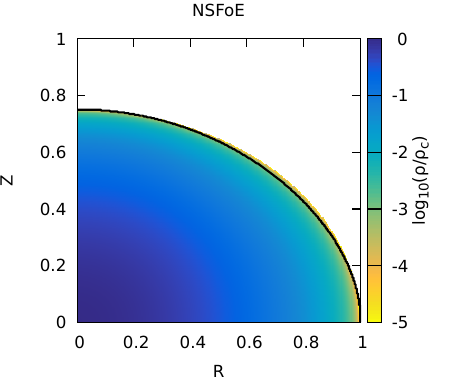}\\\medskip
    \includegraphics[height=0.6\linewidth,trim={16.75cm 0.2cm 0.55cm 0.4cm},clip]{comp1_750_15_08.pdf}
    \caption{Color-coded mass density map (log. scale) computed from Eq. \eqref{eq:eq_set_1D} that takes advantage of dimension reduction (top) and absolute difference (log. scale) with the reference {\tt DROP} code (bottom), for configuration B. The parameters are $\bar\varepsilon_{\rm s}=0.75$ and $n=1.5$,   with $257$ nodes per direction (see also Table \ref{tab:configB}).}
    \label{fig:comp1_750_15_08}
\end{figure}

\subsection{Example of two-dimensional reconstruction (deprojection)}\label{subsec:reconstruction}

Once a solution in the form of a pair of profiles $\hatrho(\varpi)$ and $\varepsilon(\varpi)$ is known, we can reconstruct the $2$D structure by using Eq. \eqref{eq:ellipse}. For this second example, we ran the SCF code under the same conditions as above, but for $\bare_{\rm s}=0.75$ and $n=1.5$, which could correspond to a fast-rotating fully convective star. We note that  $\bare_{\rm s}\approx0.74$ for Achernar \citep{dkm14}. We can compare the $\rho$-map obtained in this way to the field output by the {\tt DROP} code. This is shown in Fig. \ref{fig:comp1_750_15_08}. We see that the largest differences are mainly located close to the surface, as the  true solution is slightly sub-elliptical. The deviations do not exceed about $3\times10^{-3}$ in absolute (this value is on the order of the virial parameter, see below). This result is remarkable, in particular because the surface is the place where the mass density is small and vanishes. The best agreement is observed at the center, at the pole, and at the equator, with absolute deviations of about $10^{-5}$. The main quantities are reported in Table \ref{tab:configB}. We see that most global quantities are correctly reproduced within a percent, which is satisfying for a rotator  this fast.

\begin{table}[ht]
    \centering
    \caption{Same caption and same conditions as for Table \ref{eq:eq_set}, but for a fast rotator.}
    \begin{tabular}{lrr}
        configuration B & {\tt DROP}$^{\dagger}$ code & this work \\\hline
        $\bare_{\rm s}$ & $\leftarrow 0.750$ & $\leftarrow 0.750$ \\
        $n$ & $\leftarrow 1.5$ & $\leftarrow 1.5$ \\
        $M/(\rho_{\rm c}R_{\rm e}^3)$ & $4.3026\times10^{-1}$ & $4.3397\times10^{-1}$ \\
        $I/(\rho_{\rm c}R_{\rm e}^5)$ & $7.4701\times10^{-2}$ & $7.5961\times10^{-2}$\\
        $V/R_{\rm e}^3$ & $3.0297\textcolor{white}{xxxxxx}$ & $3.1415\textcolor{white}{xxxxxx}$\\
        $\hatomega^2$ & $2.2662\times10^{-1}$ &$2.2808\times10^{-1}$ \\
                      &                      & $^\star2.2807\times10^{-1}$\\
        $J/(G\rho_{\rm c}^3R_{\rm e}^{10})^{1/2}$ & $3.5561\times10^{-2}$ & $3.6278\times10^{-2}$ \\
        $-W/(G\rho_{\rm c}^{2}R_{\rm e}^{5})$ & $1.8344\times10^{-1}$ & $1.8584\times10^{-1}$ \\
        $T/(G\rho_{\rm c}^{2}R_{\rm e}^{5})$ & $8.4644\times10^{-3}$ & $8.6630\times10^{-3}$ \\
        $U/(G\rho_{\rm c}^{2}R_{\rm e}^{5})$ & $1.6651\times10^{-1}$ & $1.6790\times10^{-1}$\\
        $|{\rm VP}/W|$ & $1\times10^{-5}$ & $3\times10^{-3}$\\
        \hline
    \end{tabular}
    \label{tab:configB}
\end{table}

\subsection{Varying the surface axis ratio and the polytropic index: Computing vs. precision}

We   performed similar comparisons by varying the surface axis ratio $\bare_{\rm s}$ and the polytropic index $n$, again with a moderate numerical resolution corresponding to $N=256$. The results are gathered in Table \ref{tab:drop_ecs_comp} in Appendix \ref{app:tab}, where we   list  the mass, the rotation rate, the relative  virial parameter and the root mean square (RMS) difference between the two structures (mass, density). There are three series. For the first the index and the resolution are held fixed ($n=1.5$, $N=256$) and the surface eccentricity increases  to the critical rotation, at about $\bare_{\rm s} \approx 0.62$. In the second series the surface eccentricity and the resolution are fixed ($\bare_{\rm s}=0.95$, $N=256$) and $n$ varies from $0.5$ to $4$. In the last series, the configuration is fixed ($\bare_{\rm s}=0.95$, $n=1.5$) and we increase the resolution from $32$ to $2048$. We see that the agreement is good: the maximum RMS value is $4\times10^{-3}$ and the mass and rotation rate are well reproduced within a percent. From the virial parameters and the RMS, two trends are seen. First, the method is less and less accurate as the axis ratio decreases, especially for a ``hard'' EOS. This is expected because the true surface deviates   more from a perfect spheroid as the rotation becomes faster. Second, the method is increasingly accurate as the polytropic index increases (``soft'' EOS), even close to critical rotations. As $n$ increases, the density becomes peaked at the center, and the contribution to the mass (and potential) of the ``wings'' becomes small to negligible. This is  visible in Fig. \ref{fig:ecs_prog}a, for instance. For $n=3$ the mass density vanishes quickly toward the surface ($\hat\rho \lesssim 10^{-3}$ at $\varpi \gtrsim 0.7$).

The computing time reported in the table is obtained on a standard laptop, without any specification optimization. For $n=1.5$ and $\bare_{\rm s}=0.95$, the number of iterations is about $30$ and is not   sensitive to $N$. We find $\sim \left(N/581.295\right)^{1.930}$ seconds for convergence with the SSB approximation, to be compared to the full $2$D problem $\sim \left(N/77.824\right)^{3.089}$ seconds.

\section{More tests}
\label{sec:tests}

In order to better see the power of the method and its flexibility, we present in this section several tests, including static and rotating configurations (see next section for systems with mass density jumps). The convergence criterion and the numerical resolution are (unless stated otherwise) the same as in Sect. \ref{subsec:example}.

\begin{figure}
    \centering
   \includegraphics[width=\linewidth,trim={0.2cm 0.1cm 0.2cm 0.2cm},clip]{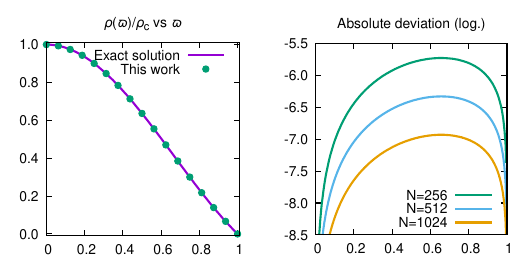}
   \caption{Mass density profile obtained by solving Eq. (\ref{eq:eq_set_1D}) with the SCF algorithm ($N=256$) in the static case ($\bare_{\rm s}=1)$ and for  $n=1$, compared to the corresponding exact solution of the Lane-Emden equation (left panel). Also shown is the absolute deviation between these two solutions, for several values of $N$ (right panel); see also Table \ref{tab:nonrota_unit}.}
    \label{fig:nonrota_unit}
\end{figure}

\subsection{Static case with index unity}

We consider here a static gas with polytropic index $n=1$ (see the Introduction). All isobars are spherical and $\varepsilon=0$ for any $\varpi$. In this case the solution of the Lane-Emden equation is analytical,\footnote{This is also the case for $n=0$ and $n=5$, but $n=0$ represents a homogeneous object and this test was already   considered in Paper IV. The case with $n=5$ has an infinite radial extent and mass; see Sect. \ref{subsec:extp}.\label{note:n}} namely
\begin{equation}\label{eq:n1_nonrot}
  \hat\rho(\varpi) = \frac{\sin(\uppi\varpi)}{\uppi\varpi},
\end{equation}
which enables a direct comparison. The theory of NSFoE is exact for spherical configuration because the confocal parameters are null, which is therefore the case of Eq. \eqref{eq:eq_set_1D}. If we inject $\varepsilon(\varpi)=0$ in the four functions $\kappa$, $\chi$, $\mu$, and $\eta$, an expansion is required in the limit $\varepsilon \rightarrow 0$. We find that $\varepsilon=0$ is in fact a regular singular point, and it follows that $\kappa=\chi=0$ (see   Appendix \ref{app:spherelim}). Thus, we recover that the body is  nonrotating and that all isopycnics are spherical. We have  applied the SSB approximation. The results are shown in Fig. \ref{fig:nonrota_unit}. The main output data are listed in Table \ref{tab:nonrota_unit}. We see that the deviation from the analytical solution is of order $10^{-5}$, and it decreases as $N$ increases. This occurs because deviations are directly linked to the order of the quadrature scheme in this case.

\begin{table}[ht]
    \centering
    \caption{Mass and moment of inertia for the static polytropic system with $n=1$ (second column), and values computed from the SCF method, using the 2D code (third column) and after dimension reduction (fourth column).}
    \begin{tabular}{lrrr}  
           & exact & {\tt DROP} code & this work \\\hline   
           $M/(\rho_{\rm c}R_{\rm e}^3)$ & $1.27324$ & $1.27322$ & $1.27323$\\
           $I/(\rho_{\rm c}R_{\rm e}^5)$ & $0.33280$ & $0.33279$ & $0.33279$\\
           Steps for convergence & & $29$ & $20$ \\
           Virial parameter & 0 & $4 \times 10^{-6}$ & $2 \times 10^{-5}$\\\hline   
    \end{tabular}
    \label{tab:nonrota_unit}
\end{table}

\subsection{The case of slow rotation}

In the limit of slow rotation, various approximations can be found \citep[see, e.g.,][]{ch33}. Of particular interest is Clairaut's theory, which is first-order accurate in the square of the eccentricity, $\varepsilon^2\ll 1$. It happens that there are a few closed-form solutions to Clairaut's second-order differential equation compatible with physically realistic BCs. Among them, \citet{tisserand91} and \citet{mar00} have a Legendre-Laplace solution for $n=1$: the solution is the same as for the nonrotating case (i.e., a sine cardinal for the mass density).\footnote{As Clairaut's equation is  first order accurate in $\varepsilon^2$, only the zeroth-order on $\hat\rho$ is needed to have $\varepsilon^2$ in the limit of slow rotations.} For Eq.\eqref{eq:n1_nonrot}, the eccentricity profile reads
\begin{equation}\label{eq:tisserand}
    \bare^2(\varpi) = 1-\left(1-\bare^2_{\rm s}\right)\frac{(\uppi^2\varpi^2-3)\sin(\uppi\varpi)+3\uppi\varpi\cos(\uppi\varpi)}{3\varpi^2\left[\uppi\varpi\cos(\uppi\varpi)-\sin(\uppi\varpi)\right]}
.\end{equation}

\begin{figure}
    \centering
    \includegraphics[width=\linewidth,trim={0.2cm 0.1cm 0.2cm 0.2cm},clip]{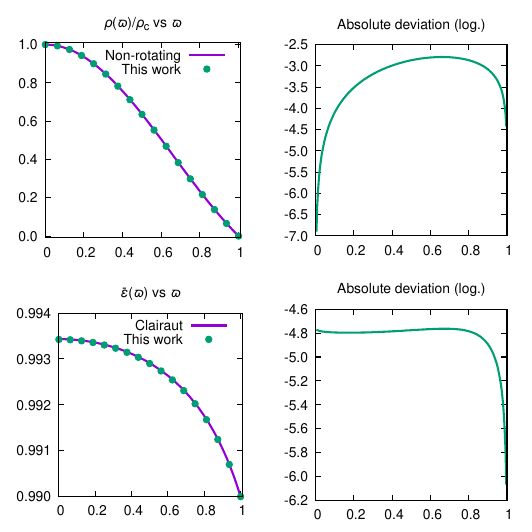}
    \caption{Mass density profile and squared eccentricity profile ({left panels}) obtained by solving Eq. \eqref{eq:eq_set_1D} with the SCF algorithm, and comparison with the exact nonrotating first-order solution from Clairaut's fundamental equation ({right panels}).}
    \label{fig:rota_unit}
\end{figure}

\begin{figure}
    \centering
    \includegraphics[height=0.6\linewidth,trim={8.2cm 0.2cm 8cm 0.4cm},clip]{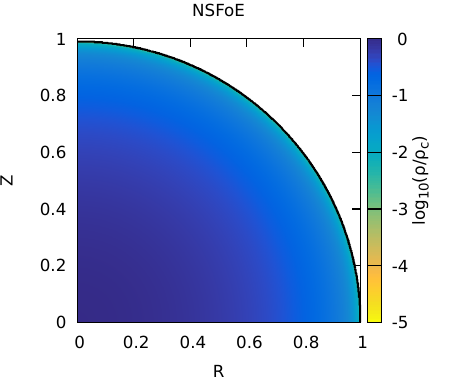}\\\medskip
    \includegraphics[height=0.6\linewidth,trim={16.75cm 0.2cm 0.55cm 0.4cm},clip]{comp1_990_10_09.pdf}
    \caption{Same as Fig. \ref{fig:comp1_750_15_08}, but for $\bar\varepsilon_{\rm s}=0.99$ and $n=1$.}
    \label{fig:comp1_990_10_08}
\end{figure}

We ran the code for $n=1$ and $\bare_{\rm s}=0.99$. The results are displayed in Fig. \ref{fig:rota_unit}. We see that the mass density profile of the rotating case departs only slightly from a sine cardinal (the deviation is of order $10^{-3}$). More importantly, the actual method yields an eccentricity profile that is very close to Eq. \eqref{eq:tisserand}, within a few $10^{-5}$ in absolute, which is highly satisfying. This confirms the efficiency of the SSB approximation at slow rotation. Figure \ref{fig:comp1_990_10_08} displays the mass density structure and the deviation from the reference structure, with a RMS value of $2\times10^{-6}$.

\begin{figure}
    \centering
    \includegraphics[height=0.6\linewidth,trim={8.2cm 0.2cm 8cm 0.4cm},clip]{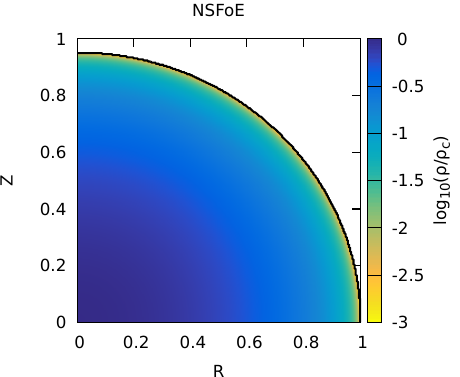}\\\medskip
    \includegraphics[height=0.6\linewidth,trim={16.75cm 0.2cm 0.55cm 0.4cm},clip]{comp1_950_10_08.pdf}
    \caption{Same  as  Fig. \ref{fig:comp1_750_15_08}, but for $\bar\varepsilon_{\rm s}=0.95$ and $n=1$.}
    \label{fig:comp1_950_10_08}
\end{figure}

\begin{figure*}
    \centering
    \includegraphics[width=0.9\linewidth]{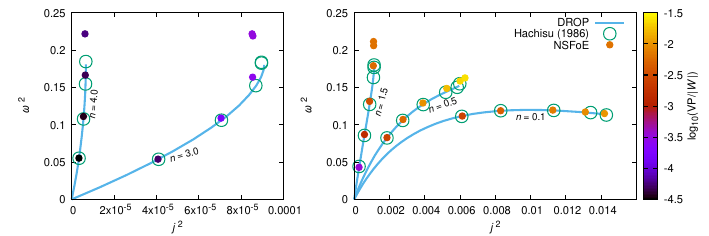}
    \caption{The $(j^2,\omega^2)$ diagram for several polytropic indices $n$ labeled on the curves. The solid lines correspond to the solutions obtained from {\tt DROP} code. The empty circles are the solutions reported in the tables of \citet{hachisu86}. The full circles are the solutions obtained with the SCF method with dimension reduction. The color-coding indicates the decimal logarithm of the relative virial parameter $|{\rm VP}/W|$.}
    \label{fig:w2j2}
\end{figure*}

\subsection{A case of moderate rotation}

An interesting test concerns the transition from slow to moderate rotators. We  performed a new run for $n=1$ and $\bare_{\rm s}=0.95$ (which is close to the axis ratio of Jupiter, for instance). The results are shown in Fig. \ref{fig:comp1_950_10_08} after reconstruction of the $2$D map for the mass density. We see that the deviation is again maximum at the surface, midway between the pole and the equator. The highest value of the RMS for the mass density is about $3\times10^{-4}$, while the mean value is about $7\times10^{-5}$. At the center, near the pole and the equator, the precision is excellent, with more than six correct digits.

\subsection{Reproduction of Hachisu's \texorpdfstring{$(j^2,\omega^2)$}{} sequences of equilibrium}

In the 1980s Hachisu and collaborators   perfomed a wide exploration of figures of equilibrium through a series of fundamental papers \citep{hachisu86,hachisu86III}. They  computed sequences of  axially symmetrical equilibria by varying the  surface axis ratio $\bare_{\rm s}$, both for stars and rings (out of range here). The configurations are gathered in the $(j^2,\omega^2)$-plane, where $\omega$ and $j$ are the normalized rotation rate $\varOmega$ and angular momentum $J$, respectively, defined as
\begin{equation}
    \left\{
        \begin{aligned}
            &j^2 = \frac{J^2}{4\uppi GM^3V^{1/3}}\\
            &\omega^2 = \frac{\varOmega^2}{4\uppi G MV^{-1}}
        \end{aligned}.
    \right.
\end{equation}
\citet{hachisu86}   showed that, in contrast to the Maclaurin uncompressible sequence, all sequences are open, depending on the polytropic index $n>0$ of the gas: the larger the value of  $n$, the larger the gap. Figure \ref{fig:w2j2} shows a few points of the sequences obtained with the SSB approximation for $n=\{0.1,0.5,1.5,3,4\}$, together with the data extracted from Hachisu's paper (same axis ratio). Values obtained in full $2$D with the {\tt DROP} code  are also plotted for comparison. We see that the agreement is excellent, except for points close to critical rotations. This is obviously due to the deviation of the external surface to a spheroid: the volume is overestimated in this case, which causes a double shift in the diagram, toward higher $\omega$-values and lower $j$-values. The best results are obtained for weakly compressible gas. The virial parameters, on the other hand, are better and better as $n$ increases (see section \ref{subsec:reconstruction}).

\section{Cases with mass density jumps}
\label{sec:jumps}

\subsection{Changes}

Bodies made of different inhomogeneous domains separated by mass density jumps are of immense interest. Such cases are studied in Paper IV. We now consider ${\cal K}$ domains where the mass density $\rho_k(\varpi)$ varies (domain no. $1$ is for the innermost domain and $k=\cal K$ is for the outermost one). Then we can write the full mass density profile from the center to the surface as
\begin{equation}\label{eq:rho_jumps}
    \hatrho(\varpi) = \sum_{k=1}^{\cal K} \left[\hatrho_k(\varpi)-\hatrho_{k+1}(\varpi)\right]{\cal H}(\varpi_k-\varpi),
\end{equation}
where $\varpi_k$ is the position, along the equatorial axis, of the mass density jump between domains $k$ and $k+1$, and $\cal H$ is Heaviside's distribution. For convenience we   set $\rho_{{\cal K}+1}=0$ to keep a single sum in Eq. \eqref{eq:rho_jumps}. The radial derivative is then given by the expression
\begin{align}\label{eq:drho_jumps}
    \frac{\du\hat\rho}{\du \varpi} = \sum_{k=1}^{\cal K} \bigg[\frac{\du\hat\rho_k}{\du \varpi}&-\frac{\du\hat\rho_{k+1}}{\du \varpi}\bigg]{\cal H}(\varpi_k-\varpi) \notag\\ &- \sum_{k=1}^{\cal K} \left[\hat\rho_k(\varpi)-\hat\rho_{k+1}(\varpi)\right]\delta(\varpi_k-\varpi),
\end{align}
where $\delta$ is Dirac's distribution. So, for any continuous function $f(\varpi',\varpi)$ in the interval $[0,1]^2$, we have
\begin{align} \notag
   & \int_{\hatrho(0)}^{\hatrho(1)} \du\hatrho(\varpi')f(\varpi',\varpi) = \sum_{k=1}^{\cal K} \int_{\hatrho_k(\varpi_{k-1})}^{\hatrho_k(\varpi_k)} \du\hatrho_k(\varpi')f(\varpi',\varpi) \\
    &\mkern+175mu -\sum_{k=1}^{\cal K} \frac{\alpha_k-1}{\alpha_k}\hatrho_k(\varpi_k) f(\varpi_k,\varpi),\label{eq:fintegral}
\end{align}
where $\varpi_0=0$ again for convenience, and the mass density jump $\alpha_k$ is defined by
\begin{equation}
    \alpha_k = \frac{\rho_k(\varpi_k)}{\rho_{k+1}(\varpi_k)}.
\end{equation}
It follows that the structure of a rigidly rotating body made of several inhomogeneous domains can be treated with dimension reduction by solving Eqs. \eqref{eq:hent} and \eqref{eq:ide_compact}, where all the integrals are estimated from Eq. \eqref{eq:fintegral}.

\subsection{Presence of an ambient pressure}
\label{subsec:extp}

\begin{figure}
    \centering
    \includegraphics[width=\linewidth,trim={0.2cm 0.1cm 0.2cm 0.2cm},clip]{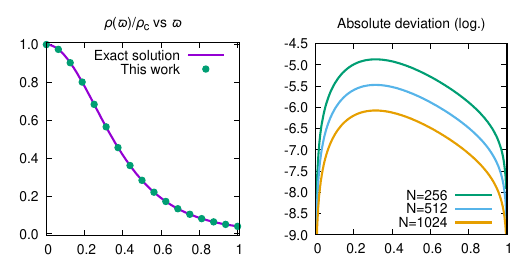}
    \caption{Mass density profile obtained by solving Eq.\eqref{eq:ide_compact} with the SCF algorithm and $N=256$, with configuration $\bare_{\rm s}=1$, $n=5$, and ${\hat\rho_{\rm amb}=0.04}$ (configuration C), compared to the exact solution of Schuster ({left panel}), and the absolute deviation between these two solutions for several values of $N$ ({right panel}).}
    \label{fig:nonrot_pamb}
\end{figure}

As the present formalism relies mainly on a barotropic EOS, we assume the external pressure to be constant along the outermost surface of the object:  $P_{\rm amb}$. Thus, this value corresponds to a cutoff for the mass density at $\hat\rho_{\rm amb}$, and then to a mass density jump at $\varpi=1$. In the case of a single domain object (${\cal K}=1$ and $\varpi_1=1$ here), we have 
\begin{equation}\label{eq:rho_extp}
    \hatrho(\varpi) = \hatrho_1(\varpi){\cal H}(1-\varpi),
\end{equation}
where $\hatrho_1(1)=\hatrho_{\rm amb}$. An interesting test case for ambient pressure is the $n=5$ polytrope. The analytical solution in the nonrotating case is due to Schuster (\citealt[e.g.,][]{horedttextbook2004}; see note \ref{note:n}). In the present context, the solution must be truncated at the right isobar $P(1)=P_{\rm amb}$. The analytical solution becomes
\begin{equation}
    \hat\rho(\varpi) = \frac{1}{[1+\varpi^2(\hat\rho_{\rm amb}^{-2/5}-1)]^{5/2}},
\end{equation}
which verifies $\hat\rho(0)=1$ and $\hat\rho(1)=\hat\rho_{\rm amb}$. We  computed the solution from the SSB approximation. We note that the method is not appropriate  for the case $\hat\rho_{\rm amb}=0$ because the mass and especially the radius are infinite for $P_{\rm amb}=0$. The reconstructed mass density map is displayed in Fig. \ref{fig:nonrot_pamb}, and the main data are listed in Table \ref{tab:nonrot_pamb}. As for the $n=1$ static case, we see that the agreement is excellent and the deviation from the exact solution depends just on the resolution, as expected. We note that the virial parameter was adapted to the context of an ambient pressure. We have 
\begin{equation}
    {\rm VP} = W + 2T + U - U_{\rm amb},
\end{equation}
where $U_{\rm amb}=3P_{\rm amb}V$ (see, e.g., \citealt{cox_1968}).

\begin{table}
    \centering
    \caption{Data associated with Fig. \ref{fig:nonrot_pamb} (static pressurized case).}
    \begin{tabular}{lrrr}  
           & exact & this work \\\hline   
           $M/\rho_{\rm c}R_{\rm e}^3$ & $0.60718$ & $0.60720$\\ 
           $I/\rho_{\rm c}R_{\rm e}^5$ & $0.15262$ & $0.15262$\\ 
           Steps for convergence & & $21$ \\
           Virial parameter & 0 & $2 \times 10^{-5}$\\\hline   
    \end{tabular}
    \label{tab:nonrot_pamb}
\end{table}

\begin{figure}
    \centering
    \includegraphics[height=0.6\linewidth,trim={8.2cm 0.2cm 8cm 0.4cm},clip]{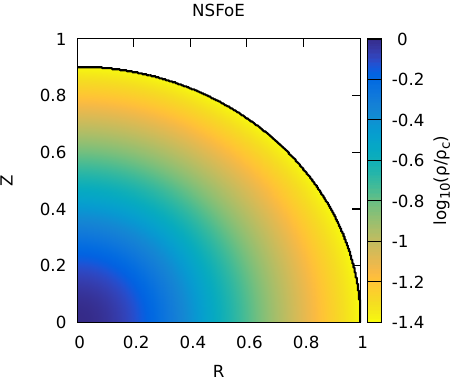}\\\medskip
    \includegraphics[height=0.6\linewidth,trim={16.75cm 0.2cm 0.55cm 0.4cm},clip]{comp1_900_50_08.pdf}
    \caption{Same  as  Fig. \ref{fig:comp1_750_15_08}, but for $\bar\varepsilon_{\rm s}=0.90$, $n=5.0$, and $\hat\rho_{\rm amb}=0.04$ (configuration C); see also Table \ref{tab:rot_pamb}.}
    \label{fig:rot_pamb}
\end{figure}

\begin{table}
    \centering
    \caption{Same caption and same conditions as for Table \ref{tab:configB}, but for the rotating case with ambient pressure.}
    \begin{tabular}{lrr}
        Configuration C & {\tt DROP}$^{\dagger}$ code & this work \\\hline
        $\bare_{\rm s}$ & $\leftarrow 0.900$ & $\leftarrow 0.900$ \\
        $n$ & $\leftarrow 5$ & $\leftarrow 5$ \\
        $\hat\rho_{\rm amb}$ & $\leftarrow 0.04$ & $\leftarrow 0.04$ \\
        $M/(\rho_{\rm c}R_{\rm e}^3)$ & $5.2781\times10^{-1}$ & $5.2871\times10^{-1}$ \\
        $I/(\rho_{\rm c}R_{\rm e}^5)$ & $1.3129\times10^{-1}$ & $1.3169\times10^{-2}$\\
        $V/R_{\rm e}^3$ & $3.7596\textcolor{white}{xxxxxx}$ & $3.7699\textcolor{white}{xxxxxx}$\\
        $\hatomega^2$ & $7.7911\times10^{-2}$ &$7.8114\times10^{-2}$ \\
                      &                      & $^\star7.8109\times10^{-2}$\\
        $J/(G\rho_{\rm c}^3R_{\rm e}^{10})^{1/2}$ & $3.6648\times10^{-2}$ & $3.6805\times10^{-2}$ \\
        $-W/(G\rho_{\rm c}^{2}R_{\rm e}^{5})$ & $2.2649\times10^{-1}$ & $2.2709\times10^{-1}$ \\
        $T/(G\rho_{\rm c}^{2}R_{\rm e}^{5})$ & $5.1147\times10^{-3}$ & $5.1434\times10^{-3}$ \\
        $U/(G\rho_{\rm c}^{2}R_{\rm e}^{5})$ & $2.7139\times10^{-1}$ & $2.7190\times10^{-1}$\\
        $U_{\rm amb}/(G\rho_{\rm c}^{2}R_{\rm e}^{5})$ & $5.5118\times10^{-2}$ & $5.5283\times10^{-2}$\\
        $|{\rm VP}/W|$ & $5\times10^{-5}$ & $8\times10^{-4}$\\
        \hline
    \end{tabular}
    \label{tab:rot_pamb}
\end{table}

\begin{figure*}
    \centering
    \includegraphics[width=\linewidth,trim={0.5cm 0.0cm 0.5cm 0.0cm},clip]{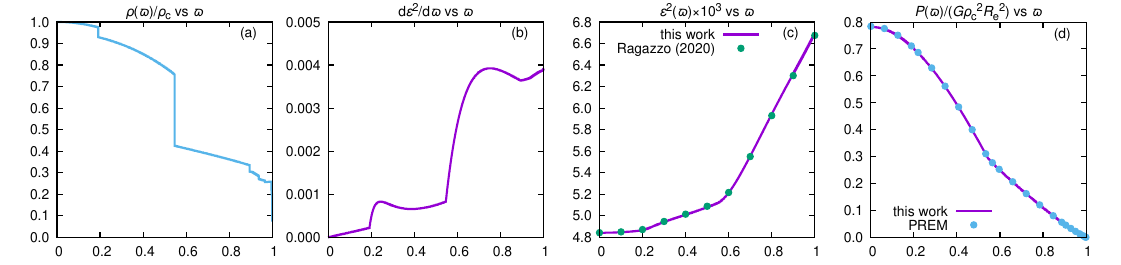}
    \caption{Main outputs for the Earth using the PREM. {From left to right}: Normalized mass density of the PREM \citep{prem81} which is used as an input in this work, and the gradient of the eccentricity squared, squared eccentricity, and normalized pressure determined from the SSB approximation.}
    \label{fig:earth_prem}
\end{figure*}

In the rotating case, Schuster's solution is not adapted anymore, and we switch back to the {\tt DROP} code for the reference. For this illustration we consider the configuration given in Table \ref{tab:rot_pamb}. It is an intermediate rotator with the same index and ambient pressure as before. A comparison between the full $2$D solution and the reconstructed density field, as well as the difference between the two maps,  is shown in Fig. \ref{fig:rot_pamb}. We see that the RMS error\footnote{As the true boundary of the fluid is slightly sub-elliptical, some nodes are out of the fluid in the $2$D problem, while $\hatrho=\hatrho_{\rm amb}$. To avoid an excessive overestimation of the RMS error, we   subtracted the ambient mass density from both maps to obtain this value.} is on the order of $10^{-4}$. The main global quantities are given in Table \ref{tab:rot_pamb}. The agreement is once again very good, with local deviations on the order of $10^{-4}$ on the density profile. For global properties, the relative deviations are on the order of $10^{-3}$.

\subsection{The Earth as a an example}
\label{subsec:earth}

\citet{prem81}   obtained a seismologic model for the internal structure of the Earth, named the Preliminary Reference Earth Model (PREM). They derived profiles for the mass density, pressure, and gravitational field of the Earth under the hypothesis of a purely spherical planet. In this section we use the mass density profile of the PREM as an input to model a rotating Earth with the SSB approximation; it is plotted in Fig. \ref{fig:earth_prem}a. It is clear that nonspherical effects have an impact on the mass density distribution. However, the Earth is a slow rotator, so the deviation is expected to be on the order of $1-\bare_{\rm s}\sim 3\times10^{-3}$ in relative. The PREM is a model with ${\cal K}=10$ domains, and as many mass density jumps.\footnote{As the atmosphere is not   taken into account in the PREM, a mass density jump is present at the outermost surface: from liquid water to the outer space.} As $\hat\rho(\varpi)$ is an input, the SSB approximation developed in this work returns to a single fixed-point problem where only step (i) of the cycle is needed (i.e., we iterate only on Eq. \eqref{eq:ide_compact}). Only $15$ iterations are required by the SCF method (with $N=1024$ per domain). We show the results for $\du\varepsilon^2/\du\varpi$ and $\varepsilon^2(\varpi)$ in Fig. \ref{fig:earth_prem}b and c, respectively. We   also report  the eccentricity profile computed by \citet{ragazzo2020}, who   solved Clairaut's equation with the same piece-wise prescription for the mass density.
We see that both solutions are in excellent agreement. The pressure profile $P(\varpi)$ at equilibrium is visible in Fig. \ref{fig:earth_prem}d (see Appendix \ref{app:pres_proof} for this calculation). It compares very well to the PREM values (the central dimensionless value is $\sim 0.782$ comparted to $0.784$ for the PREM). The relative deviations do not exceed $\sim 3 \times10^{-3}$, which is on the order of the flattening of the outermost surface. This departure is not due to the spheroidal approximation made in this work (as $|{\rm VP}/W| \sim 7 \times 10^{-7}$), but to the fact that the observed mass density profile has been spherically averaged by the PREM and that centrifugal effects have not been taken into account in their calculations of the pressure profile.

The global properties deduced from SSB approximation are reported in Table \ref{tab:earth}. However, the data in this table must be rescaled prior to any comparison with observational data. For this purpose we chose to use the mass and equatorial radius given by \citet{crv10}, and reported in Table \ref{tab:earth_units}. We then deduced $\rho_{\rm c}=13083.8~{\rm kg\cdot m^{-3}}$, whose value remains close to the PREM value of  $1.3088\times10^{4}~{\rm kg\cdot m^{-3}}$. Table \ref{tab:earth_units} compares the momentum of inertia, rotation rate, and first two gravitational moments. We see that the normalized moment of inertia and the rotation rates are in good agreement with the observational data, again within $\sim 3 \times10^{-3}$ in relative. Regarding the gravitational moments, we see that $J_2$ is close to the observed value, roughly within $1 \%$. The result is worse for $J_4$, a deviation reaching $75\%$. A similar discrepancy is found by \citet{crv10}, who   solved the second-order Clairaut's equation with a different (yet similar) mass density profile for the Earth. As quoted by these authors, this deviation from the observed gravitational moments is due to nonhydrostatic characteristics of the Earth, which are not taken into account in our equation set.

\begin{table}
    \centering
    \caption{Output dimensionless quantities obtained for the rotating Earth from the SSB approximation, for $N=10240$ in total.}
    \begin{tabular}{lr}    
           & \multicolumn{1}{c}{this work} \\\hline   
           $\bare_{\rm s}$ & $\leftarrow 0.99665$\\
           $M/(\rho_{\rm c}R_{\rm e}^3)$ & $1.7592\textcolor{white}{xxxxxx}$\\
           $I/(\rho_{\rm c}R_{\rm e}^5)$ & $5.8188\times10^{-1}$\\
           $\hatomega^2$ & $6.1199\times10^{-3}$ \\
           $\hat{P}_{\rm c}$ & $7.8200\times10^{-1}$ \\
           $J/(G\rho_{\rm c}^3R_{\rm e}^{10})^{1/2}$ & $4.5521\times10^{-2}$\\
           $-W/(G\rho_{\rm c}^2R_{\rm e}^5)$ & $2.0631\textcolor{white}{xxxxxx}$\\
           $T/(G\rho_{\rm c}^2R_{\rm e}^5)$ & $1.7805\times10^{-3}$\\
           $U/(G\rho_{\rm c}^2R_{\rm e}^5)$ & $2.0596\textcolor{white}{xxxxxx}$\\
           $|{\rm VP}/W|$ & $7\times10^{-7}$\\\hline   
    \end{tabular}
    \label{tab:earth}
\end{table}

\begin{table}
    \centering
    \caption{Physical properties of the Earth rescaled; $R_{\rm v}=R_{\rm e}\bare_{\rm s}^{1/3}$ is the mean volumetric radius ($ \approx 6.371\times10^{6}$ m).}
    \begin{tabular}{lrrr}
           &  &   \multicolumn{1}{c}{second-order}&  \\
          data & \multicolumn{1}{c}{observed$^{\dagger}$} &  \multicolumn{1}{c}{Clairaut$^{\dagger}$}& \multicolumn{1}{c}{this work} \\\hline
          $M~\rm [kg]$ & \multicolumn{1}{c}{$5.97218\times10^{24}$}\\
          $R_{\rm e}~\rm [m]$ & \multicolumn{1}{c}{$6.378137\times10^{6}$}\\\\
          $I/MR_{\rm v}^2$              & $3.3069\times10^{-1}$ &                       & $3.3151\times10^{-1}$ \\
          $\varOmega~\rm [s^{-1}]$     & $7.2921\times10^{-5}$ &                       & $7.3104\times10^{-5}$ \\
          $J_2$                        & $1.0826\times10^{-3}$ & $1.0712\times10^{-3}$ & $1.0771\times10^{-3}$ \\
          $-J_4$                       & $1.620\times10^{-6}$ & $2.96\times10^{-6}$    & $2.8233\times10^{-6}$ \\\hline
          \multicolumn{4}{l}{$^{\dagger}$see \citet{crv10} and references therein.}
    \end{tabular}
    \label{tab:earth_units}
\end{table}

\section{Summary}
\label{sec:conclusion}

In this article  we   showed that the $2$D structure of a rotating self-gravitating fluid can be determined with   good precision from the theory of Nested Spheroidal Figures of Equilibrium, which assumes that isopycnics are perfect spheroids. The method uses the general IDE established in \cite{sh24}, connecting the local eccentricity $\varepsilon$ of isopynics to the local mass density $\rho$. As this equation involves only the equatorial radius, it can be coupled to the centrifugal balance along the polar axis, providing a specific one-dimensional projection of the genuine bi-dimensional problem. As shown, the new problem consists in solving two fixed-points problems coupled together. From the equatorial solution $\{\rho(a),\varepsilon(a)\}$, which can be determined by a simple SCF method, we  recover the full structure by propagating the mass density along the ellipse, from the equator to the pole, and for any equatorial radius $a \in [0,\re]$. We   provided a series of examples supporting the efficiency and versatility of this SSB approximation. In particular, the method is well suited to systems made of heterogeneous domains separated by mass density jumps. It can account for an ambient pressure. The method is also flexible in terms of barotropic EOS. Here we   used a polytrope, but any kind of $P(\rho)$-relation is usable. The main results of this paper can be summarized as follows:
\begin{itemize}
\item The SSB approximation is exact for nonrotating gas, because the NSFoE is also exact in this case (all confocal parameters are null). The method is then equivalent to a standard Lane-Emden solver. The precision is then fully governed by the numerical schemes implemented.  
\item For slow rotators (i.e., small flattenings), the method has an excellent precision. Depending on the EOS, the solution can reproduce the full $2$D mass density profile, typically within $10^{-5}$ in absolute (dimensionless profile), even at low to moderate numerical resolution.
\item For fast rotators (i.e., large flattenings), the SSB approximation furnishes good and reliable results, whatever the EOS.
\item For hard EOSs (typically $0<n<1$), the deviation of the true surface from a spheroid is very small, but there is a significant amount of matter close to the surface. The precision of SSB approximation is very good provided the system stays far from the critical rotation state. Near the sequence endings, the precision is acceptable, with typically $1 \%$ in the mass density (RMS value).
\item For soft EOSs ($n>1$) the precision is very good because the amount of matter located close to the surface has a negligible contribution to the total mass (and gravitational potential), although the deviation of the true surface from a spheroid is significant.
\item Compared to the full $2$D problem in which the surface and isoycnics are not constrained, the mass density in the vicinity of the center computed with the SSB approximation has an excellent precision. This is also true for values close to the pole and to the equator.
\item The computing time is reduced by at least two orders of magnitude with respect to the full $2$D problem. This is a direct consequence of dimension reduction. This enables us to reach very high numerical resolutions in a short time, which is particularly attractive for generating grids of models for instance.
\end{itemize}

\begin{acknowledgements}
    We are grateful to F. Chambat for informed comments on the paper before submission. We thank the anonymous referee for suggestions to improve the paper and valuable references, in particular on the state-of-art regarding stellar models.
\end{acknowledgements}

\bibliographystyle{aa}

\begin{thebibliography}{57}
    \expandafter\ifx\csname natexlab\endcsname\relax\def\natexlab#1{#1}\fi
    
    \bibitem[{{Abramyan} \& {Kaplan}(1974)}]{ak74}
    {Abramyan}, M.~G. \& {Kaplan}, S.~A. 1974, Astrophysics, 10, 358
    
    \bibitem[{{Amendt} {et~al.}(1989){Amendt}, {Lanza}, \&
      {Abramowicz}}]{amendt1989}
    {Amendt}, P., {Lanza}, A., \& {Abramowicz}, M.~A. 1989, \apj, 343, 437
    
    \bibitem[{{Basillais} \& {Hur{\'e}}(2021)}]{bh21}
    {Basillais}, B. \& {Hur{\'e}}, J.-M. 2021, \mnras, 506, 3773
    
    \bibitem[{{Binney} \& {Tremaine}(1987)}]{binneytremaine87}
    {Binney}, J. \& {Tremaine}, S. 1987, {Galactic dynamics} (Princeton, NJ,
      Princeton University Press, 1987, 747 p.)
    
    \bibitem[{{Chambat} {et~al.}(2010){Chambat}, {Ricard}, \& {Valette}}]{crv10}
    {Chambat}, F., {Ricard}, Y., \& {Valette}, B. 2010, Geophysical Journal
      International, 183, 727
    
    \bibitem[{{Chandrasekhar}(1933)}]{ch33}
    {Chandrasekhar}, S. 1933, \mnras, 93, 390
    
    \bibitem[{{Chandrasekhar}(1969)}]{chandra69}
    {Chandrasekhar}, S. 1969, {Ellipsoidal figures of equilibrium} (Yale Univ.
      Press)
    
    \bibitem[{{Chandrasekhar} \& {Roberts}(1963)}]{cr63}
    {Chandrasekhar}, S. \& {Roberts}, P.~H. 1963, \apj, 138, 801
    
    \bibitem[{{Cisneros Parra} {et~al.}(2015){Cisneros Parra}, {Mart{\'\i}nez
      Herrera}, \& {Montalvo Castro}}]{cmm15}
    {Cisneros Parra}, J.~U., {Mart{\'\i}nez Herrera}, F.~J., \& {Montalvo Castro},
      J.~D. 2015, \rmxaa, 51, 121
    
    \bibitem[{Cox \& Giuli(1968)}]{cox_1968}
    Cox, J.~P. \& Giuli, R.~T. 1968, Principles of stellar structure {Volume} {I} :
      {Physical} principles, 1st edn. (Gordon and Breach)
    
    \bibitem[{{Debras} \& {Chabrier}(2018)}]{dc18}
    {Debras}, F. \& {Chabrier}, G. 2018, \aap, 609, A97
    
    \bibitem[{{Domiciano de Souza} {et~al.}(2014){Domiciano de Souza}, {Kervella},
      {Moser Faes}, {Dalla Vedova}, {M{\'e}rand}, {Le Bouquin}, {Espinosa Lara},
      {Rieutord}, {Bendjoya}, {Carciofi}, {Hadjara}, {Millour}, \&
      {Vakili}}]{dkm14}
    {Domiciano de Souza}, A., {Kervella}, P., {Moser Faes}, D., {et~al.} 2014,
      \aap, 569, A10
    
    \bibitem[{Dziewonski \& Anderson(1981)}]{prem81}
    Dziewonski, A.~M. \& Anderson, D.~L. 1981, Physics of the Earth and Planetary
      Interiors, 25, 297
    
    \bibitem[{{Espinosa Lara} \& {Rieutord}(2013)}]{elr13}
    {Espinosa Lara}, F. \& {Rieutord}, M. 2013, \aap, 552, A35
    
    \bibitem[{{Even} \& {Tohline}(2009)}]{et09}
    {Even}, W. \& {Tohline}, J.~E. 2009, \apjs, 184, 248
    
    \bibitem[{Gradshteyn \& Ryzhik(2014)}]{gradshteyn15}
    Gradshteyn, I.~S. \& Ryzhik, I.~M. 2014, in Table of Integrals, Series, and
      Products, 8th edn., ed. D.~Zwillinger \& V.~Moll (Boston: Academic Press),
      776
    
    \bibitem[{{Hachisu}(1986{\natexlab{a}})}]{hachisu86}
    {Hachisu}, I. 1986{\natexlab{a}}, \apjs, 61, 479
    
    \bibitem[{{Hachisu}(1986{\natexlab{b}})}]{hachisu86III}
    {Hachisu}, I. 1986{\natexlab{b}}, \apjs, 62, 461
    
    \bibitem[{{Hadjifotinou}(2000)}]{had00}
    {Hadjifotinou}, K.~G. 2000, \aap, 354, 328
    
    \bibitem[{{Horedt}(2004)}]{horedttextbook2004}
    {Horedt}, G.~P., ed. 2004, Astrophysics and Space Science Library, Vol. 306,
      {Polytropes - Applications in Astrophysics and Related Fields}
    
    \bibitem[{{Houdayer} \& {Reese}(2023)}]{hr23}
    {Houdayer}, P.~S. \& {Reese}, D.~R. 2023, \aap, 675, A181
    
    \bibitem[{{Hubbard}(2013)}]{hub13}
    {Hubbard}, W.~B. 2013, \apj, 768, 43
    
    \bibitem[{{Hur{\'e}}(2022{\natexlab{a}})}]{h2022a}
    {Hur{\'e}}, J.-M. 2022{\natexlab{a}}, \mnras, 512, 4031
    
    \bibitem[{{Hur{\'e}}(2022{\natexlab{b}})}]{h2022b}
    {Hur{\'e}}, J.-M. 2022{\natexlab{b}}, \mnras, 512, 4047
    
    \bibitem[{{Hur{\'e}} \& {Hersant}(2017)}]{hh17}
    {Hur{\'e}}, J.-M. \& {Hersant}, F. 2017, \mnras, 464, 4761
    
    \bibitem[{{Hur{\'e}} {et~al.}(2018){Hur{\'e}}, {Hersant}, \& {Nasello}}]{hhn18}
    {Hur{\'e}}, J.-M., {Hersant}, F., \& {Nasello}, G. 2018, \mnras, 475, 63
    
    \bibitem[{{James}(1964)}]{j64}
    {James}, R.~A. 1964, \apj, 140, 552
    
    \bibitem[{{Kadam} {et~al.}(2016){Kadam}, {Motl}, {Frank}, {Clayton}, \&
      {Marcello}}]{ka16}
    {Kadam}, K., {Motl}, P.~M., {Frank}, J., {Clayton}, G.~C., \& {Marcello}, D.~C.
      2016, \mnras, 462, 2237
    
    \bibitem[{{Kiuchi} {et~al.}(2010){Kiuchi}, {Nagakura}, \& {Yamada}}]{kiu10}
    {Kiuchi}, K., {Nagakura}, H., \& {Yamada}, S. 2010, \apj, 717, 666
    
    \bibitem[{{Klu{\'z}niak} \& {Rosi{\'n}ska}(2013)}]{kr13}
    {Klu{\'z}niak}, W. \& {Rosi{\'n}ska}, D. 2013, \mnras, 434, 2825
    
    \bibitem[{{Kovetz}(1968)}]{kov68}
    {Kovetz}, A. 1968, \apj, 154, 999
    
    \bibitem[{{Lander} \& {Jones}(2009)}]{lj09}
    {Lander}, S.~K. \& {Jones}, D.~I. 2009, \mnras, 395, 2162
    
    \bibitem[{{Liu}(1996)}]{liu96}
    {Liu}, F.~K. 1996, \mnras, 281, 1197
    
    \bibitem[{{Mach}(2012)}]{mach12}
    {Mach}, P. 2012, Journal of Mathematical Physics, 53, 062503
    
    \bibitem[{{Marchenko}(2000)}]{mar00}
    {Marchenko}, A.~N. 2000, Astronomical School's Report, 1, 34
    
    \bibitem[{{Mishra} \& {Vaidya}(2015)}]{mv15}
    {Mishra}, B. \& {Vaidya}, B. 2015, \mnras, 447, 1154
    
    \bibitem[{{Montalvo} {et~al.}(1983){Montalvo}, {Martinez}, \&
      {Cisneros}}]{mmc83}
    {Montalvo}, D., {Martinez}, F.~J., \& {Cisneros}, J. 1983, \rmxaa, 5, 293
    
    \bibitem[{{Nettelmann}(2017)}]{net17}
    {Nettelmann}, N. 2017, \aap, 606, A139
    
    \bibitem[{{Odrzywo{\l}ek}(2003)}]{od03}
    {Odrzywo{\l}ek}, A. 2003, \mnras, 345, 497
    
    \bibitem[{{Ostriker} \& {Mark}(1968)}]{om68}
    {Ostriker}, J.~P. \& {Mark}, J.~W.-K. 1968, \apj, 151, 1075
    
    \bibitem[{Ragazzo(2020)}]{ragazzo2020}
    Ragazzo, C. 2020, São Paulo Journal of Mathematical Sciences, 14, 1
    
    \bibitem[{{Rampalli} {et~al.}(2023){Rampalli}, {Smock}, {Newton}, {Daniel}, \&
      {Curtis}}]{ramp23}
    {Rampalli}, R., {Smock}, A., {Newton}, E.~R., {Daniel}, K.~J., \& {Curtis},
      J.~L. 2023, \apj, 958, 76
    
    \bibitem[{{Rieutord}(2006)}]{r06}
    {Rieutord}, M. 2006, in SF2A-2006: Semaine de l'Astrophysique Francaise, ed.
      D.~{Barret}, F.~{Casoli}, G.~{Lagache}, A.~{Lecavelier}, \& L.~{Pagani}, 501
    
    \bibitem[{{Rieutord} {et~al.}(2016){Rieutord}, {Espinosa Lara}, \&
      {Putigny}}]{rep16}
    {Rieutord}, M., {Espinosa Lara}, F., \& {Putigny}, B. 2016, Journal of
      Computational Physics, 318, 277
    
    \bibitem[{{Roberts}(1963)}]{rob63}
    {Roberts}, P.~H. 1963, \apj, 138, 809
    
    \bibitem[{{Roxburgh} \& {Strittmatter}(1966)}]{rs66}
    {Roxburgh}, I.~W. \& {Strittmatter}, P.~A. 1966, \mnras, 133, 345
    
    \bibitem[{{Seidov} \& {Kuzakhmedov}(1978)}]{sei78}
    {Seidov}, Z.~F. \& {Kuzakhmedov}, R.~K. 1978, \sovast, 22, 711
    
    \bibitem[{{Sharma}(1977)}]{sha77}
    {Sharma}, V.~D. 1977, Physics Letters A, 60, 381
    
    \bibitem[{{Srivastava}(1962)}]{sri62}
    {Srivastava}, S. 1962, \apj, 136, 680
    
    \bibitem[{{Staelen} \& {Hur{\'e}}(2024)}]{sh24}
    {Staelen}, C. \& {Hur{\'e}}, J.-M. 2024, \mnras, 527, 863
    
    \bibitem[{Tassoul(1978)}]{tassoul78}
    Tassoul, J.-L. 1978, Theory of Rotating Stars. (Princeton University Press)
    
    \bibitem[{Tisserand(1891)}]{tisserand91}
    Tisserand, F. 1891, Trait{\'e} de m{\'e}canique c{\'e}leste - II. Th{\'e}orie
      de la figure des corps c{\'e}lestes et de leur mouvement de rotation
      (Gauthier-Villars et fils)
    
    \bibitem[{{Tohline}(2021)}]{tohlinewiki21}
    {Tohline}, J.~E. 2021, a (MediaWiki-based) Vistrails.org publication,
      https://www.vistrails.org/index.php/User:Tohline
    
    \bibitem[{{Tomimura} \& {Eriguchi}(2005)}]{te05}
    {Tomimura}, Y. \& {Eriguchi}, Y. 2005, \mnras, 359, 1117
    
    \bibitem[{{Venditti} {et~al.}(2020){Venditti}, {de Almeida Junior}, \&
      {Prado}}]{vdap20}
    {Venditti}, F. C.~F., {de Almeida Junior}, A.~K., \& {Prado}, A. F.~B.~A. 2020,
      \planss, 192, 105063
    
    \bibitem[{V\'eronet(1912)}]{veronet12}
    V\'eronet, A. 1912, Journal de math{\'e}matiques pures et appliqu{\'e}es 6e
      s{\'e}rie, 8, 331
    
    \bibitem[{{Zharkov} \& {Trubitsyn}(1970)}]{zt70}
    {Zharkov}, V.~N. \& {Trubitsyn}, V.~P. 1970, \sovast, 13, 981
    
    \end{thebibliography}

\appendix
\onecolumn

\section{The functions \texorpdfstring{$\chi$}{}, \texorpdfstring{$\mu$}{}, \texorpdfstring{$\nu$}{}, and \texorpdfstring{$\eta$}{}}
\label{app:chimunueta}

We define for convenience
\begin{equation}
  e^2(c)=\frac{\varpi'^2\varepsilon^2(\varpi')}{\varpi^2[1+c]},
\end{equation}
and $\bar{e}(c)=\sqrt{1-e^2(c)}\ge0$, where the continuous confocal parameter $c \equiv c(\varpi',\varpi)$ is given by 
\begin{equation}
    c(\varpi',\varpi) = \frac{\varpi'^2\varepsilon^2(\varpi')}{\varpi^2}-\varepsilon^2(\varpi).
\end{equation}In these conditions, and given the function ${\cal A}(\bare)$ defined by Eq.\eqref{eq:mla}, $\chi$ is defined as 
\begin{equation}\label{eq:chi_def}
    \chi(\varpi',\varpi;\varepsilon) = \left\{
        \begin{aligned}
            &\frac{\varpi'^3\bare(\varpi')}{\varpi^4}\left[{\cal A}\big(\bar{e}(0)\big)-\frac{{\cal A}\big(\bar{e}(c)\big)}{\sqrt{1+c}}\right], &\varpi'<\varpi\\
            &0, &\varpi'\geq\varpi
        \end{aligned},
        \right.
\end{equation}
$\mu$ is defined as 
\begin{equation}\label{eq:mu_def}
    \mu(\varpi',\varpi;\varepsilon) = \left\{
    \begin{aligned}
        &\frac{\varpi'\bare(\varpi'){\cal A}\big(\bar{e}(c)\big)}{\varpi\varepsilon^2(\varpi')\sqrt{1+c}} - \frac{\varpi'\bare(\varpi')}{\varpi\varepsilon^2(\varpi')\bare(\varpi)}, &\varpi'<\varpi\\
        &\frac{\bare(\varpi')}{\varepsilon^2(\varpi')}{\cal A}\big(\bare(\varpi')\big)-\frac{1}{\varepsilon^2(\varpi')}, &\varpi'\geq\varpi
    \end{aligned},
    \right.
\end{equation}
$\eta$ is defined as 
\begin{equation}\label{eq:eta_def}
    \eta(\varpi',\varpi;\varepsilon) = \left\{
    \begin{aligned}
        &\varpi'\varpi\frac{\bare(\varpi')}{\varepsilon^2(\varpi')}\left[{\cal A}\big(\bar{e}(c)\big)\sqrt{1+c}-\bare(\varpi)\right] &\varpi'<\varpi\\
        &\varpi^2[1+c]\frac{\bare(\varpi')}{\varepsilon^2(\varpi')}{\cal A}\big(\bare(\varpi')\big) - \varpi^2\frac{\bare^2(\varpi)}{\varepsilon^2(\varpi')}, &\varpi'\geq\varpi
    \end{aligned},
    \right.
\end{equation}
and $\kappa$ is defined as
\begin{equation}\label{eq:kappa_def}
    \kappa(\varpi',\varpi;\varepsilon) = \left\{
    \begin{aligned}
        &\frac{\varpi'\bare(\varpi')}{\varpi\varepsilon^2(\varpi')}\Bigg\{\left[1-2e^2(0)\right]{\cal A}\big(\bar{e}(0)\big)-2\bare(\varpi)-\bar{e}(0)+2{\cal A}\big(\bar{e}(c)\big)\sqrt{1+c}\Bigg\}, &\varpi'<\varpi\\
        &\frac{3-2\varepsilon^2(\varpi)}{\varepsilon^2(\varpi')}\left[\bare(\varpi'){\cal A}\big(\bare(\varpi')\big)-1\right]+1, &\varpi'\geq\varpi
    \end{aligned}.
    \right.
\end{equation}

\section{A single function for the prolate and oblate cases}
\label{app:onefunction}

As can be seen in Paper IV, which is  devoted to oblate configurations, all functions $\kappa$, $\chi$, $\mu$, (and $\eta$) depend on $\varepsilon$, which appears in the argument of an $\arcsin$ function. During the numerical tests, we dealth with transition states where the deep isopycnic surfaces are slightly prolate, resulting in a purely imaginary eccentricity. This naturally creates numerical issues, which can be circumvented by a simple prolongation since $\arcsin(z) = {\rm arcsinh}({\rm i}z)/{\rm i}$, where $\rm i$ is the imaginary unit and $z$ is a complex number \citep[see, e.g.,][]{gradshteyn15}. So, for the numerical computation we define
\begin{equation}\label{eq:mla}
    {\cal A}(\bare) = \left\{
        \begin{aligned}
            &\frac{\arcsin(\sqrt{1-\bare^2})}{\sqrt{1-\bare^2}}, &0<\bare<1,\\
            &1, &\bare=1,\\
            &\frac{{\rm arcsinh}(\sqrt{\bare^2-1})}{\sqrt{\bare^2-1}}, &\bare>1,\\
        \end{aligned}
    \right.
\end{equation}
which enables us to make the transition from oblate to prolate configurations. It can be verified (see Appendix \ref{app:chimunueta}), by introducing this function $\cal A$ in all integrand kernels defined in Paper IV, that only real terms like $\bare$ or $\varepsilon^2$ finally survive, whatever the case, either oblate or prolate.

\section{The enthalpy profile}
\label{app:hent_proof}

Along the polar axis $R=0$, the Bernoulli equation Eq. \eqref{eq:bernoulli} reads
\begin{equation}\label{eq:bernoulli_polar}
    H(0,Z) + \varPsi(0,Z) = \const.
\end{equation}In order to estimate $\varPsi(0,Z)$, we need to go back to Paper II. The gravitational potential of a heterogeneous system made of $\cal L$ homogeneous spheroidal layers is given in that article by Eq.  (14). At point ${\rm A}_j$, at the intersection between the polar axis and the interface between layers $j$ and $j+1$, it reads
\begin{align}
    \frac{\varPsi|_{{\rm A}_j}}{-2\uppi G a_j^2} = &\sum_{i=1}^{j} (\rho_i-\rho_{i+1})\frac{\bare_i}{\varepsilon_i^3}\left[(1+c_{i,j})\arcsin\left(\frac{q_{i,j}\varepsilon_i}{\sqrt{1+c_{i,j}}}\right)-q_{i,j}\varepsilon_i\bare_j\right]\notag\\
    &\mkern+200mu+\sum_{i=1}^{j} (\rho_i-\rho_{i+1})\frac{\bare_i}{\varepsilon_i^3}\left[(1+c_{i,j})\arcsin\left(\varepsilon_i\right)\left(q_{i,j}^2\varepsilon_i^2+\bare_j^2\right)-\frac{\varepsilon_i\bare_j^2}{\bare_i^2}\right],
\end{align}
where $a_j$ is the equatorial radius of layer $j$, $\varepsilon_j$ is its eccentricity, $\rho_j$ is its mass density, $\bare_j=\sqrt{1-\varepsilon_j^2}$, $q_{i,j}=a_i/a_j$, and $c_{i,j}=q_{i,j}^2\varepsilon_i^2-\varepsilon_j^2$. Oblate spheroids were assumed in Paper II, but it can be generalized to any spheroid by introducing the function $\cal A$ defined in Appendix \ref{app:onefunction}. In the continuous limit ${\cal L}\rightarrow \infty$, we have
\begin{equation}
    \frac{\varPsi(R=0,Z)}{G\rho_{\rm c}\re^2} = 2\uppi\int_0^1 \du \hat\rho(\varpi') \eta(\varpi',\varpi;\varepsilon)
.\end{equation}

As $H(0,Z)$ is a function of $\varpi$ only, we define $\hat{H}(\varpi)=H(0,Z)/(G\rho_{\rm c}\re^2)$. By Eq.\eqref{eq:bernoulli_polar}, we have
\begin{equation}\label{eq:hent_app}
    \hat{H}(\varpi) + 2\uppi\int_0^1 \du \hat\rho(\varpi') \eta(\varpi',\varpi;\varepsilon) = \hat{H}(1) + 2\uppi\int_0^1 \du \hat\rho(\varpi') \eta(\varpi',1;\varepsilon),
\end{equation}
where the constant $C^{\rm te}$ was evaluated at the pole of the whole object (i.e., at $\varpi=1$). We then see that Eq. \eqref{eq:hent_app} thus yields Eq. \eqref{eq:hent}.

\section{The spherical limit}
\label{app:spherelim}

For $\varepsilon^2\ll 1$, we have
\begin{equation}
    {\cal A}(\bare) \approx 1 + \frac{\varepsilon^2}{6}
\end{equation}
for both $\bare<1$ and $\bare>1$. So we can expand functions $\kappa$, $\chi$, $\mu$, and $\eta$ at zeroth order to obtain these functions at $\bare=1$ (i.e., in the nonrotating spherical case). We have $\kappa(\varpi',\varpi) = 0$, which naturally imposes $\hatomega^2=0$ (see Eq. \eqref{eq:rrate}), which means that the body is nonrotating, as expected. Regarding the functions $\chi$ and $\mu$ appearing in the expression for $\du\varepsilon^2/\du\varpi$, we have $\chi(\varpi',\varpi) = 0$ and
\begin{equation}
    \mu(\varpi',\varpi) = \left\{
        \begin{aligned}
            & -\frac{\varpi'^3}{3\varpi^3}, &\varpi'<\varpi\\
            & -\frac13, &\varpi'\geq\varpi
        \end{aligned}.
    \right.
\end{equation}
This thus imposes $\du\varepsilon^2/\du\varpi=0$ with BC $\varepsilon^2(1)=0$. We then recover $\varepsilon^2(\varpi)=0$ (i.e., all isopycnic surfaces are spherical, as expected). Finally, the function $\eta$ needed to compute the enthalpy field becomes
\begin{equation}
    \eta(\varpi',\varpi) = \left\{
        \begin{aligned}
            & \frac{2\varpi'^3}{3\varpi}, &\varpi'<\varpi\\
            & \varpi'^2 - \frac{\varpi^2}{3}, &\varpi'\geq\varpi
        \end{aligned},
    \right.
\end{equation}

\section{Pressure from the enthalpy field}
\label{app:pres_proof}

For any isentropic barotrope, the enthalpy and the pressure are related by 
\begin{equation}
    \du \hat{H} = \du \hat{P}/\hatrho,
\end{equation}
where $\hat{P} = P/(G\rho_{\rm c}^2\re^2)$ is the dimensionless pressure. So, the pressure profile is given by 
\begin{equation}
    \hat{P}(\varpi) = \hat{P}(1) - \int_{\varpi}^1 \du\varpi'\hatrho(\varpi')\frac{\du\hat{H}}{\du\varpi'}
.\end{equation}We can compute the enthalpy gradient from Eq.\eqref{eq:hent}. We can show that 
\begin{equation}
    \frac{\du\hat{H}}{\du\varpi} = -2\uppi \left[2\varpi\bare^2(\varpi)-\varpi^2\frac{\du\varepsilon^2}{\du\varpi^2}\right]\int_{\hatrho(0)}^{\hatrho(1)} \du\hatrho(\varpi')\mu(\varpi',\varpi).
\end{equation}

\newpage

\section{Varying the surface axis ratio and the polytropic index: Computing versus precision}
\label{app:tab}

\begin{table*}[ht]
    \centering
    \caption{Data obtained from dimension reduction compared to values obtained for the full $2$D problem with the {\tt DROP} code, obtained for several surface axis ratios $\bare_{\rm s}$ and polytropic indices $n$. The computing times spent to achieve convergence are purely indicative, and may vary a little  with the computer load.}
    \begin{tabular}{lccccccccc}
    & $N$         & $n$   & $\bare_{\rm s}$ & iterations & comp. time (s) & $M/(\rho_{\rm c}R_{\rm e}^3)$ & $\varOmega^2/(G\rho_{\rm c})$ & $|{\rm VP}/W|$ & RMS\\\hline
  this work       & $256$ & $1.5$     & $0.950$   & $32 $   & $0.199$  & $6.492\times10^{-1}$ & $5.314\times10^{-2}$ & $1\times10^{-4}$ & $4\times10^{-5}$\\
  {\tt DROP code} &       &           &           & $32 $   & $37.0$   & $6.490\times10^{-1}$ & $5.310\times10^{-2}$ & $1\times10^{-5}$ & \\
                  & $256$ & $1.5$     & $0.900$   & $33 $   & $0.217$  & $5.979\times10^{-1}$ & $1.034\times10^{-1}$ & $5\times10^{-4}$ & $2\times10^{-4}$\\
                  &       &           &           & $44 $   & $48.4$   & $5.973\times10^{-1}$ & $1.032\times10^{-1}$ & $1\times10^{-5}$ & \\
                  & $256$ & $1.5$     & $0.850$   & $35 $   & $0.230$  & $5.452\times10^{-1}$ & $1.501\times10^{-1}$ & $1\times10^{-3}$ & $4\times10^{-4}$\\
                  &       &           &           & $42 $   & $46.6$   & $5.437\times10^{-1}$ & $1.497\times10^{-1}$ & $1\times10^{-5}$ & \\
                  & $256$ & $1.5$     & $0.800$   & $37 $   & $0.249$  & $4.906\times10^{-1}$ & $1.921\times10^{-1}$ & $2\times10^{-3}$ & $7\times10^{-4}$\\
                  &       &           &           & $42 $   & $46.1$   & $4.881\times10^{-1}$ & $1.912\times10^{-1}$ & $1\times10^{-5}$ & \\
                  & $256$ & $1.5$     & $0.750$   & $39 $   & $0.261$  & $4.339\times10^{-1}$ & $2.280\times10^{-1}$ & $3\times10^{-3}$ & $1\times10^{-3}$\\
                  &       &           &           & $44 $   & $48.0$   & $4.302\times10^{-1}$ & $2.266\times10^{-1}$ & $1\times10^{-5}$ & \\
                  & $256$ & $1.5$     & $0.700$   & $43 $   & $0.285$  & $3.747\times10^{-1}$ & $2.561\times10^{-1}$ & $5\times10^{-3}$ & $2\times10^{-3}$\\
                  &       &           &           & $56 $   & $59.7$   & $3.698\times10^{-1}$ & $2.538\times10^{-1}$ & $2\times10^{-5}$ & \\
                  & $256$ & $1.5$     & $0.650$   & $50 $   & $0.327$  & $3.126\times10^{-1}$ & $2.736\times10^{-1}$ & $6\times10^{-3}$ & $2\times10^{-3}$\\
                  &       &           &           & $56 $   & $58.2$   & $3.071\times10^{-1}$ & $2.704\times10^{-1}$ & $2\times10^{-5}$ & \\
                  & $256$ & $1.5$     & $0.617$   & $50 $   & $0.324$  & $2.701\times10^{-1}$ & $2.777\times10^{-1}$ & $6\times10^{-3}$ & $2\times10^{-3}$\\
                  &       &           &           & $74 $   & $73.3$   & $2.648\times10^{-1}$ & $2.740\times10^{-1}$ & $2\times10^{-5}$ & \\\hline
  this work       & $256$ & $0.5$     & $0.950$   & $16 $   & $0.102$  & $2.162\textcolor{white}{xxxxxx}$ & $1.189\times10^{-1}$ & $5\times10^{-4}$ & $7\times10^{-4}$\\
  {\tt DROP code} &       &           &           & $23 $   & $28.5$   & $2.162\textcolor{white}{xxxxxx}$ & $1.189\times10^{-1}$ & $3\times10^{-4}$ & \\
                  & $256$ & $1.0$     & $0.950$   & $24 $   & $0.152$  & $1.197\textcolor{white}{xxxxxx}$ & $8.259\times10^{-2}$ & $1\times10^{-4}$ & $7\times10^{-5}$\\
                  &       &           &           & $51 $   & $56.4$   & $1.197\textcolor{white}{xxxxxx}$ & $8.253\times10^{-1}$ & $2\times10^{-5}$ & \\
                  & $256$ & $2.0$     & $0.950$   & $40 $   & $0.244$  & $3.359\times10^{-1}$ & $3.089\times10^{-2}$ & $1\times10^{-4}$ & $3\times10^{-5}$\\
                  &       &           &           & $44 $   & $47.3$   & $3.358\times10^{-1}$ & $3.087\times10^{-2}$ & $1\times10^{-5}$ & \\
                  & $256$ & $2.5$     & $0.950$   & $51 $   & $0.306$  & $1.610\times10^{-1}$ & $1.589\times10^{-2}$ & $8\times10^{-5}$ & $4\times10^{-5}$\\
                  &       &           &           & $50 $   & $53.5$   & $1.610\times10^{-1}$ & $1.589\times10^{-2}$ & $3\times10^{-5}$ & \\
                  & $256$ & $3.0$     & $0.950$   & $70 $   & $0.425$  & $6.859\times10^{-2}$ & $7.032\times10^{-3}$ & $8\times10^{-5}$ & $5\times10^{-5}$\\
                  &       &           &           & $65 $   & $68.1$   & $6.853\times10^{-2}$ & $7.023\times10^{-3}$ & $4\times10^{-5}$ & \\
                  & $256$ & $3.5$     & $0.950$   & $88 $   & $0.516$  & $2.402\times10^{-2}$ & $2.507\times10^{-3}$ & $1\times10^{-4}$ & $1\times10^{-4}$\\
                  &       &           &           & $88 $   & $91.0$   & $2.398\times10^{-2}$ & $2.501\times10^{-3}$ & $7\times10^{-5}$ & \\
                  & $256$ & $4.0$     & $0.950$   & $127$   & $0.750$  & $5.861\times10^{-3}$ & $6.163\times10^{-4}$ & $4\times10^{-4}$ & $2\times10^{-4}$\\
                  &       &           &           & $128$   & $130 $   & $5.818\times10^{-3}$ & $6.112\times10^{-4}$ & $2\times10^{-4}$ & \\\hline
  this work       & $32$  & $1.5$     & $0.950$   & $31 $   & $0.004$  & $6.505\times10^{-1}$ & $5.336\times10^{-2}$ & $5\times10^{-4}$ & $5\times10^{-4}$\\
  {\tt DROP code} &       &           &           & $34 $   & $0.067$  & $6.483\times10^{-1}$ & $5.307\times10^{-2}$ & $8\times10^{-4}$ & \\
                  & $64$  & $1.5$     & $0.950$   & $31 $   & $0.014$  & $6.496\times10^{-1}$ & $5.320\times10^{-2}$ & $2\times10^{-4}$ & $1\times10^{-4}$\\
                  &       &           &           & $34 $   & $0.497$  & $6.489\times10^{-1}$ & $5.308\times10^{-2}$ & $2\times10^{-4}$ & \\
                  & $128$ & $1.5$     & $0.950$   & $31 $   & $0.050$  & $6.493\times10^{-1}$ & $5.315\times10^{-2}$ & $1\times10^{-4}$ & $6\times10^{-5}$\\
                  &       &           &           & $35 $   & $5.10$   & $6.490\times10^{-1}$ & $5.310\times10^{-2}$ & $5\times10^{-5}$ & \\
                  & $512$ & $1.5$     & $0.950$   & $31 $   & $0.783$  & $6.493\times10^{-1}$ & $5.314\times10^{-2}$ & $1\times10^{-4}$ & $4\times10^{-5}$\\
                  &       &           &           & $33 $   & $346$    & $6.490\times10^{-1}$ & $5.311\times10^{-2}$ & $3\times10^{-6}$ & \\
   this work                 & $1024$ & $1.5$     & $0.950$   & $31 $   & $2.97$  & $6.492\times10^{-1}$ & $5.314\times10^{-2}$ & $1\times10^{-4}$ & \\
   this work                & $2048$ & $1.5$     & $0.950$   & $31 $   & $11.9$  & $6.492\times10^{-1}$ & $5.314\times10^{-2}$ & $1\times10^{-4}$ & \\\hline
   \end{tabular}
    \label{tab:drop_ecs_comp}
\end{table*}

\end{document}